\documentclass[aps, prl, twocolumn, a4paper, superscriptaddress]{revtex4-2}

\usepackage{graphicx} 
\usepackage{dcolumn} 
\usepackage{amsmath, bm} 
\usepackage{hyperref} 
\usepackage{mathrsfs} 
\usepackage{color} 
\usepackage{gensymb} 
\usepackage{braket}
\usepackage[utf8]{inputenc}
\usepackage{tikz}
\usepackage{lipsum}
\usepackage{makecell}
\usepackage{multirow}

\makeatletter
\g@addto@macro\bfseries{\boldmath}
\makeatother

\def\mns{$N_{1/3}$NbS$_2$ ($N$ = Ti, V, Mn, Fe, Co, Ni)}
\def\mnscr{$N_{1/3}$NbS$_2$ ($N$ = Ti, V, Cr, Mn, Fe, Co, Ni)}
\def\cns{Cr$_{1/3}$NbS$_2$}
\def\cts{Cr$_{1/3}$TaS$_2$}

\begin{document}

\title{Band-filling-controlled magnetism from transition metal intercalation in $N_{1/3}$NbS$_2$ revealed with first-principles calculations}

\author{Z.~Hawkhead}
\affiliation{Department of Physics, Centre for Materials Physics, Durham University, Durham, DH1 3LE, United Kingdom}
\author{T.~J.~Hicken}
\affiliation{Department of Physics, Centre for Materials Physics, Durham University, Durham, DH1 3LE, United Kingdom}
\affiliation{Department of Physics, Royal Holloway, University of London, Egham, TW20 0EX, United Kingdom}
\author{N.~P.~Bentley}
\affiliation{Department of Physics, Centre for Materials Physics, Durham University, Durham, DH1 3LE, United Kingdom}
\author{B.~M.~Huddart}
\affiliation{Department of Physics, Centre for Materials Physics, Durham University, Durham, DH1 3LE, United Kingdom}
\affiliation{Oxford University Department of Physics, Clarendon Laboratory, Parks Road, Oxford OX1 3PU, United Kingdom}
\author{S.~J.~Clark}
\affiliation{Department of Physics, Centre for Materials Physics, Durham University, Durham, DH1 3LE, United Kingdom}
\author{T.~Lancaster}
\affiliation{Department of Physics, Centre for Materials Physics, Durham University, Durham, DH1 3LE, United Kingdom}

\begin{abstract}
  We present a first-principles study of the effect of 3$d$ transition metal intercalation on the magnetic properties of the 2H-NbS$_2$ system, using spin-resolved density functional theory calculations to investigate  the electronic structure of \mnscr.
  We are able to accurately determine the magnetic moments and crystal field splitting, and find that the magnetic properties of the materials are determined by a mechanism based on filling rigid bands with electrons from the intercalant.
  We predict the dominant magnetic interaction of these materials by considering Fermi surface nesting, finding agreement with experiment where data are available.
\end{abstract}

\maketitle

The transition-metal dichalcogenides (TMDCs) [$MX_2$, where $M$ is a transition metal and $X$ is a chalcogen]~\cite{manzeli20172d,wei2018various,wang2012electronics,mak2010atomically} have earned attention over the decades for their interesting physical properties such as charge density waves~\cite{wilson1969transition,moncton1975study,rajora1987preparation,naito1982electrical}, superconductivity~\cite{tissen2013pressure,heil2017origin,witteveen2021polytypism}, thickness dependent transport properties~\cite{castellanos2012laser,ganatra2014few}, and tunable band gaps~\cite{ramasubramaniam2011tunable}.
Recently they have garnered interest for potential applications in novel low-dimensional devices such as atomically thin transistors~\cite{wang2012electronics}.
To further modify the properties, the layered van der Waals structure allows for the intercalation of other species between the TMDC layers.
In this Letter, we present spin-resolved density-functional theory (spin-DFT) calculations that elucidate the effects of intercalating the TMDCs with first period 3$d$-transition metal atoms in the form $N_{1/3}M$S$_2$, where $N$ is the intercalant.
Broadly in line with a rigid-band picture~\cite{battaglia2007non,clark1976structural}, we find that the spin-resolved electronic bands remain quite static across the series, and that the ordered spin-magnetic moment is determined by filling these crystal-field-split bands with electrons from the intercalant.
The resulting Fermi surface topology provides an explanation for the difference in dominant magnetic interactions seen in the series.
Finally, DoS analysis shows that the pseudogapped structure giving rise to the low-temperature properties of the $N$~=~Cr material~\cite{hicken2022energy,ghimire2013magnetic} should not be expected to occur near the Fermi level in other members of the series, implying that their low-temperature dynamics will be significantly different.
Where experimental comparisons are available, we find our results agree favourably in all cases.

\begin{figure}
  \centering
  \includegraphics[width=1\linewidth]{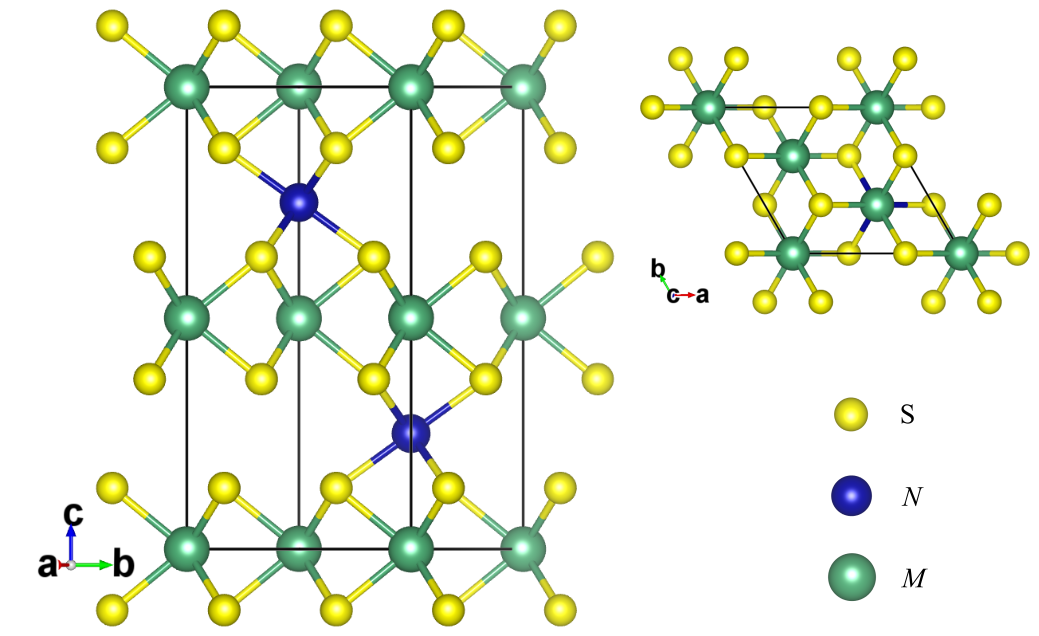}
  \caption{Crystal structure of the intercalated TMDCs. The chemical form is $N_{1/3}M$S$_2$ where $N$ are the intercalated atoms.}
  \label{fig:tmdc_structure}
\end{figure}

The structure of $N_{1/3}M$S$_2$ is shown in Fig.~\ref{fig:tmdc_structure}.
Intercalating first-period transition metals into NbS$_2$ gives ten possible materials. 
Some of these have been synthesised and widely studied; others subjected to relatively little attention, in part due to the difficult synthesis. 
Determination of the magnetic structure has been at the forefront of the effort to measure their properties, with structures so far determined for $N$~=~V~\cite{parkin19803i,parkin19803ii,hall2021magnetic}, Cr~\cite{moriya1982evidence}, Mn~\cite{kousaka2009chiral,hall2022comparative}, Fe~\cite{van1971magnetic,little2020three,xie2022structure}, Co,~\cite{parkin1983magnetic,tenasini2020giant} and Ni~\cite{battaglia2007non}.
The systems can be grouped into those hosting a ferromagnetic ($N$~=~V, Mn and Cr) or antiferromagnetic ($N$~=~Co, Ni, Fe) exchange interaction. 
Notably, amongst intercalants in the first half of the period, a ferromagnetic interaction is preferred, whereas in the second half of the period, it is more commonly antiferromagnetic. 
Of the intercalated TMDCs, the most well-studied is \cns, which hosts a long-wavelength chiral helimagnetic ground state~\cite{moriya1982evidence,miyadai1983magnetic,kousaka2016long} with near-ferromagnetic alignment of spins, and exotic magnetic textures in applied magnetic field.
Moreover, the discovery of a topologically non-trival chiral soliton lattice (regions of helimagnetism separated by ferromagnetism) on application of a magnetic field, first in \cns~\cite{kousaka2009chiral,kousaka2016long,hall2022comparative,hicken2022energy,xie2022structure,togawa2012chiral} and later in \cts~\cite{zhang2021chiral}, has renewed interest in this material class.

\begin{figure*}
  \centering
  \includegraphics[width=\linewidth]{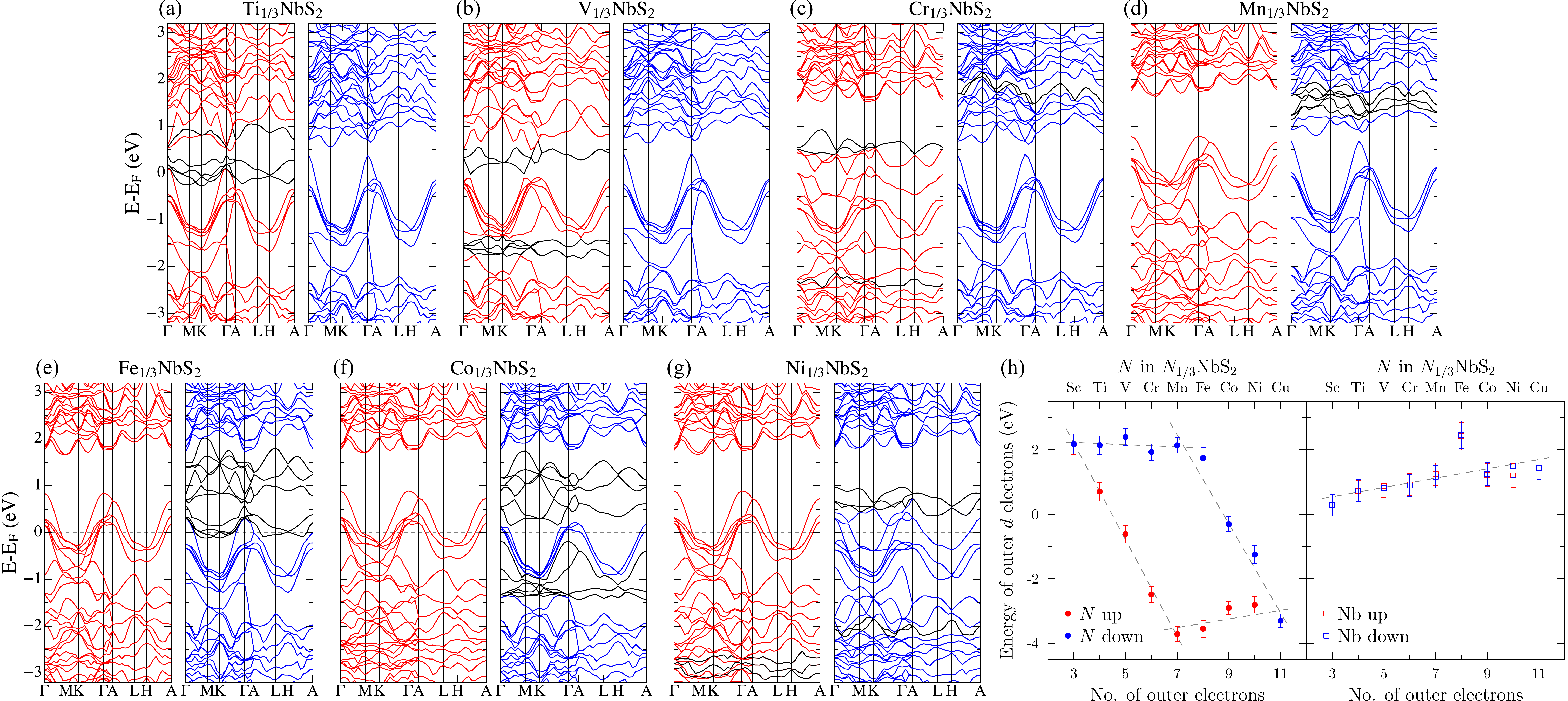}
  \caption{(a-g) Band structures of the TMDCs as a function of intercalated transition metal calculated along a labelled high-symmetry path. The band structures arranged in order of increasing unpaired electrons in the transition metal outer shell. Spin-up channels are shown in red and spin-down channels are shown in blue. The highlighted bands (black) correspond to orbitals with majority $d$ character associated with the intercalant. (h) The average band energy at the $\Gamma$-point weighted by the proportion of $N$ $d$-orbital character as a function of number of electrons in the outer shell of the intercalant. Gray dashed lines are a guide to the eye.}
  \label{fig:tmdc_bands}
\end{figure*}

Density functional theory (DFT) techniques have become increasingly useful for the study of long-range chiral magnetism over recent years~\cite{hicken2021megahertz,hicken2022magnetism}.
It was originally suggested that the photoemission and optical properties of the TMDC materials could be described using a rigid-band approximation of the electronic structure, in which intercalation leads to a transfer of charge
between the intercalant species and the $d$ conduction band of the host lattice, while the shape and ordering of the valence bands remains approximately the same \cite{battaglia2007non,clark1976structural}.
More recent first-principles calculations using spin-DFT concentrating on the $N$~=~Cr material~\cite{ghimire2013magnetic,hicken2022energy} highlight a drop in the density of states (DoS) at the Fermi level that contributes to the transport and magnetic properties at low temperatures.

We performed spin-DFT calculations of the electronic structure of $N_{1/3}$NbS$_2$, where $N$ is a transition metal in the first period, using the plane-wave, pseudopotential code, \textsc{Castep}~\cite{clark2005first,yates2007spectral}.
To more directly compare the electronic structure across the series, each calculation was initialized in a ferromagnetic state with a moment on the $N$ ions. 
Despite experimental evidence that some of these materials exhibit antiferromagnetic order in their ground state, we are only able to stabilize ferromagnetically-ordered states, suggesting that this may be a stable higher-energy configuration for some of the systems. 
To capture some of the correlation effects that often contribute to long range magnetic order, we included a small Hubbard $U$ of 2.0--2.5~eV (3.5~eV for $N$~=~Ni) on the $N$ $d$-orbitals for most materials.
These values were chosen as the smallest value required to form a ferromagnetic state in each material (below this $U$ the calculations typically result in a non-magnetic state).
It was not possible to stabilize an ordered moment at any physically-plausible value of $U$ for $N$~=~Sc, Cu, and Zn.

\begin{figure}
  \centering
   \includegraphics[width=0.99\linewidth]{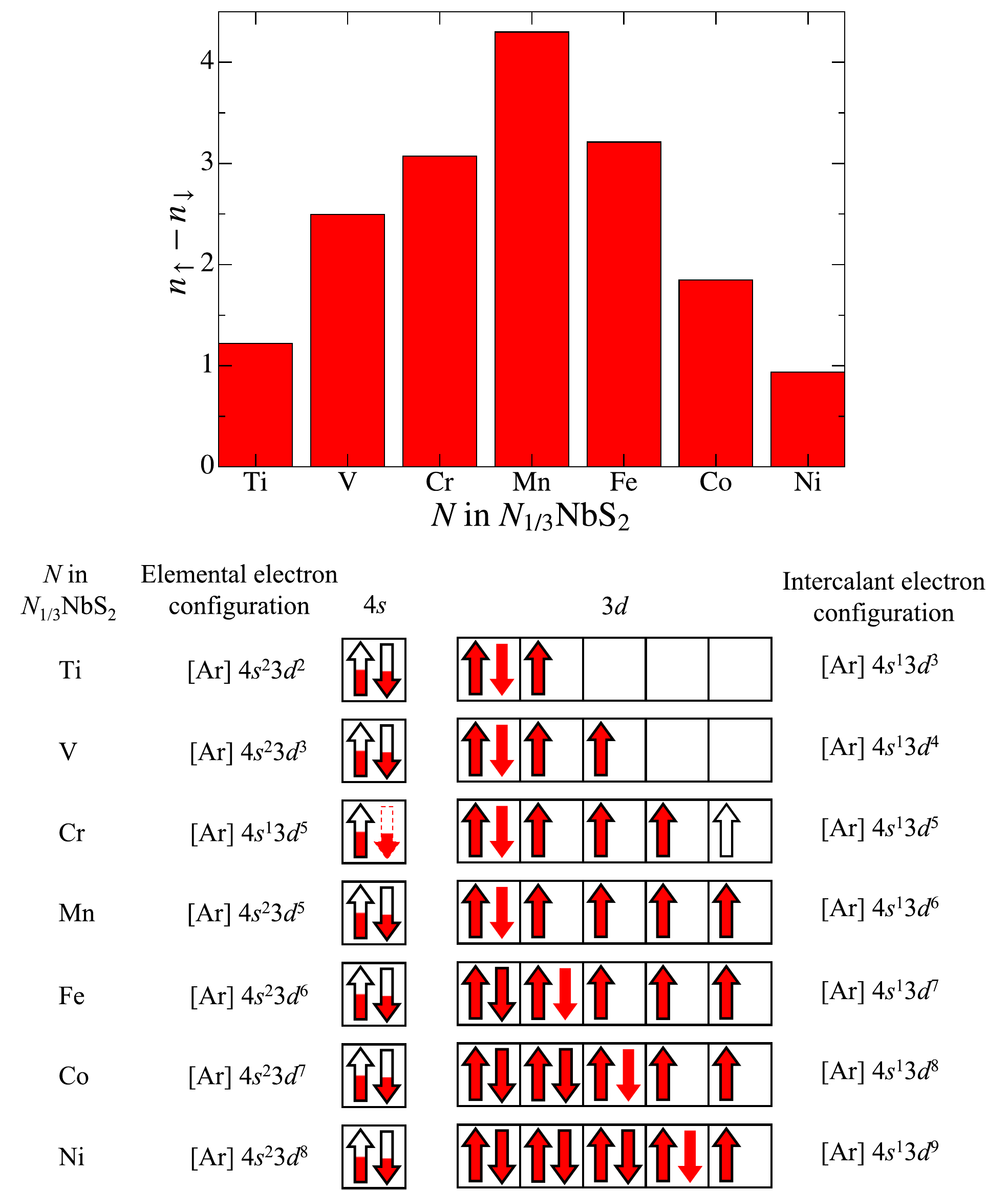}
  \caption{Top: spin-up ($n^\uparrow$) minus spin-down ($n^\downarrow$) electrons in 3$d$ orbitals of $N$ in $N_{1/3}$NbS$_2$.
  Bottom: Schematic of the band filling mechanism in the 4$s$ and 3$d$ orbitals of the intercalated transition metal compared to the elemental electronic configurations. Red arrows indicate the calculated electron occupation while black outlines show the elemental occupation. There is some hybridisation of the 4$s$ and 4$p$ orbitals in each material, with $\approx$1 electron shared across both spin-up and spin-down orbitals. The 4$p$ orbitals are omitted for clarity.}
  \label{fig:tmdc_3d}
\end{figure}

Our results imply that the magnetic properties of the intercalated TMDCs follow a simple band-filling mechanism similar to the rigid-band approximation~\cite{clark1976structural}. 
The band structures of $N_{1/3}$NbS$_2$ are shown in Fig.~\ref{fig:tmdc_bands}(a-g).
We note that as we progress from Ti$_{1/3}$NbS$_2$ through to Mn$_{1/3}$NbS$_2$, (top Fig.~\ref{fig:tmdc_bands}) the band structure of the spin-down electrons (blue) remains essentially static while the position of some spin-up electron bands shift with respect to the Fermi energy ($E_\mathrm{F}$).
Conversely, the effect is seen with the spin-species reversed for Fe$_{1/3}$NbS$_2$ through to Ni$_{1/3}$NbS$_2$, although this is less prominent for the latter materials. 
The changes to the bands contributing to the magnetic moment are characterized in Fig.~\ref{fig:tmdc_bands}(h). 
We show how a characteristic energy of the $N$ 3$d$ bands (calculated as a weighted average of the energy of the d-bands at the $\Gamma$-point) changes as the number of outer electrons is increased, demonstrating the band shifting discussed above.
In the first half of the period, the energy of the $N$ 3$d$ spin-down bands remains essentially static while the $N$ 3$d$ spin-up bands decrease in energy.
In the second half of the period, the $N$ 3$d$ spin-down band energies decrease while the spin-up band energies increase slightly.
In contrast, we see that the characteristic energy of the Nb bands [Fig.~\ref{fig:tmdc_bands}(h) (right)] is essentially static as we increase the number of electrons, with little energy difference between the spin-up and spin-down band energies.
This highlights that the magnetic behaviors in these materials are dominated by the intercalant.
We can see in Fig.~\ref{fig:tmdc_bands}(h) that the two distinct regions in the spin-up and spin-down energies intersect for Sc$_{1/3}$NbS$_2$ and Cu$_{1/3}$NbS$_2$ such that the spin-up and spin-down energy levels are equal, consistent with the lack of an overall moment found in these materials.

The static nature of a subset of the bands in the band structure has consequences for the spin found on the magnetic ions [Fig.~\ref{fig:tmdc_3d}(top)].
By projecting the electron density onto the atomic orbitals by Mulliken analysis~\cite{mulliken1955electronic}, we see that for the materials where the intercalant has a less-than-half-filled 3$d$-shell, the spin increases across the series, and then begins to drop again as we move to the materials with a more than half-filled shell.
In the first half of the period, we add approximately one electron to the spin-up subshells.
Once the spin-up shells are filled (at $N$~=~Mn) we start filling the spin-down shells, explaining the decrease in spin as we approach the end of the series.
We also see that there is just less than one electron populating the spin-down shells of the magnetic ions in the first half of the period (0.81, 0.70, 0.97, and 0.67 for $N$~=~Ti, V, Cr, and Mn respectively).
This is not what one would expect from a simple Hund's rules analysis and points to crystal field effects (see below). 
The resulting picture, illustrated in Fig.~\ref{fig:tmdc_3d}(bottom), can be extrapolated to demonstrate that the $N$~=~Sc, Cu, and Zn systems will not have any unpaired spins in the 3$d$ orbitals, explaining why they are found not to host an ordered moment (see Tab.~\ref{tab:magnetic_moment}).
Notably, we also find that the electron filling in $N$~=~Cr differs from that found in elemental Cr. 
The elemental transition metals in the first period have a similar electronic structure, with the 4$s$ shell filled and each having an extra 3$d$ electron than the previous element. 
However, this is not the case for elemental Cr, where it is energetically favorable to instead half-fill the 3$d$ band, resulting in a composition of [Ar]4$s^\mathrm{1}$3$d^\mathrm{5}$. 
Instead, \cns\ follows the pattern of the other materials in the series, consistent with a significant crystal field splitting. .

The space group for the intercalated TMDCs is $P$6$_{3}$22 and the intercalants occupy the 2c Wyckoff site, which has local site symmetry of 3.2 (corresponding to the D$_{3}$ point group). 
This environment splits the intercalant levels into $a_{1}+2e$, where the irreducible representation $a_{1}$ corresponds to the $d_{z^2}$ orbital, and $2e$ corresponds to the $d_{xz}$, $d_{yz}$, $d_{xy}$, and $d_{x^2-y^2}$ orbitals.
The overlap of the intercalant $3d$ orbitals and the sulphur $p$ orbitals results in the $d_{z^2}$ orbital being lower in energy than the other orbitals. 
By calculating the PDoS for this series, we find that the $d_{z^2}$ orbitals of the intercalant is indeed lower in energy then the other $d$ orbitals, with the majority of the DoS associated with this orbital sitting approximately 1.5~eV below the Fermi energy in these materials (see Ref.~\cite{sm}).
Capturing and quantifying these important crystal field effects increases confidence that our calculations are a faithful reflection of the underlying electronic structure.  

\begin{table}
    \begin{tabular}{c|cc|c||c|cc|c}
      & \multicolumn{2}{c}{Model} \vline & Experiment & & \multicolumn{2}{c}{Model} \vline & Experiment \\
     $M$ & $n$ & $\mu_\text{s}$ ($\mu_\text{B}$) & $\mu_\text{exp}$ ($\mu_\text{B}$) & $M$ & $n$ & $\mu_\text{s}$ ($\mu_\text{B}$) & $\mu_\text{exp}$ ($\mu_\text{B}$) \\
     \hline
    Sc & 0 & 0 & - & Fe & 3 & 3.87 & 4.8(2) \\
    Ti & 1 & 1.73 & 1.76 & Co & 2 & 2.83 & 3.0(5) \\
    V & 2 & 2.83 & 2.93(3) & Ni & 1 & 1.73 & 2.4(3) \\
    Cr & 3 & 3.87 & 3.6(2) & Cu & 0 & 0 & -\\
    Mn & 4 & 4.90 & 5.1(2) & Zn & 0 & 0 & -
    \end{tabular}
    \caption{The magnetic moment of $N_{1/3}$NbS$_2$. Calculations are from the idealised model shown in Fig.~\ref{fig:tmdc_3d}. $n$ is the number of unpaired electrons in the 3$d$ orbitals, such that the spin-only moment $\mu_\text{s}=\sqrt{n(n+2)}$. Experimental moments are an average (with a standard error when more than one value is available) of data from Refs.~\cite{van1968structural,anzenhofer1970crystal,hulliger1970magnetic,van1971magnetic,friend1977electrical,beal1979first,parkin19803d,gorochov1981transport,moriya1982evidence,haley2020half,hall2021magnetic,hall2022comparative}.}
    \label{tab:magnetic_moment}
\end{table}

We compare our calculations to experimentally measured magnetic moments by computing the spin-only magnetic moment.
\citet{Beal1979} suggested that these materials are orbitally quenched, hence the spin-only moment is a good description of the realised magnetic moment.
In Tab.~\ref{tab:magnetic_moment} we show the spin-only moment calculated using the model filling pattern shown in Fig.~\ref{fig:tmdc_3d}(bottom).
(A similar comparison from the Mulliken spin can be found in Ref.~\cite{sm}.)
We find good agreement between the calculated spin-only moments and the experimental values, with greater deviation where there is a wider range of moments reported experimentally.

\begin{figure}
  \centering
  \includegraphics[width=0.85\linewidth]{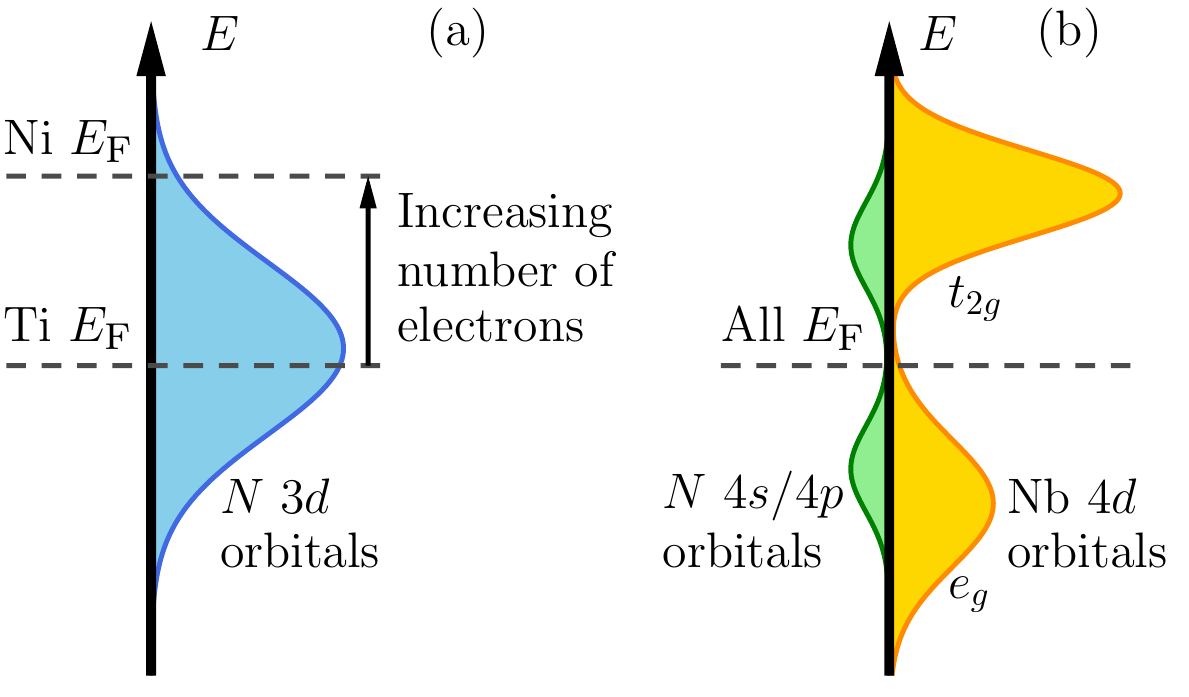}
  \caption{Schematic diagram demonstrating how the Fermi energy ($E_\mathrm{F}$) changes with respect to the non-spin-resolved DoS in \mns. (a) DoS of $N$ 3$d$ electrons, the DoS stays roughly static while $E_\mathrm{F}$ increases with added electrons. (b) DoS of $N$ 4$s$/$p$ and Nb 4$d$ electrons, the DoS stay constant with respect to $E_\mathrm{F}$.}
  \label{fig:tmdc_cartoon}
\end{figure}

The calculated DoS of \mnscr\ show marked similarities in either the spin-down ($N$~=~Ti, V, Cr, and Mn) or spin-up channels ($N$~=~Fe, Co and Ni)~\cite{sm}.
We illustrate the effects in Fig.~\ref{fig:tmdc_cartoon}.
The changes to the DoS are dominated by the $N$ 3$d$ bands.
As we increase the number of electrons by including heavier intercalants, we see that $E_\mathrm{F}$ increases relative to the $N$ 3$d$ partial DoS. 
However, the non-magnetic bands comprising the $N$ 4$s$/$p$ electrons stay fixed with respect to the changing $E_\mathrm{F}$. 
The Nb 4$d$ orbitals also undergo splitting due to the presence of crystal fields, with the $e_g$ states below $E_\mathrm{F}$ and the $t_{2g}$ states above. 
There are approximately 4 filled 4$d$ states while 6 states remain unfilled. 
We find that there is little DoS at $E_\mathrm{F}$ due to the Nb 4$d$ states.
\citet{xie2022structure} argue from symmetry considerations that there should be  a partially filled Nb 4$d_{z^2}$ orbital, however this is inconsistent with our first-principles calculations.

Our previous study of \cns\ found a depression in the DoS at $E_\mathrm{F}$ in one spin channel~\cite{hicken2022energy}.
This pseudogap is responsible for unusual low-temperature dynamics. 
However, our present results show as a result of the band-filling mechanism, this psuedogap is not found at $E_\mathrm{F}$ for any of the other materials considered.
We conclude that, rather than being a feature of the class of materials, intercalation with Cr is required for pseudogap-driven physics.
This seems to be borne out by recent experiments on the $N$~=~V and Mn systems~\cite{hall2021magnetic,hall2022comparative}.

\begin{figure}
  \centering
  \includegraphics[width=\linewidth]{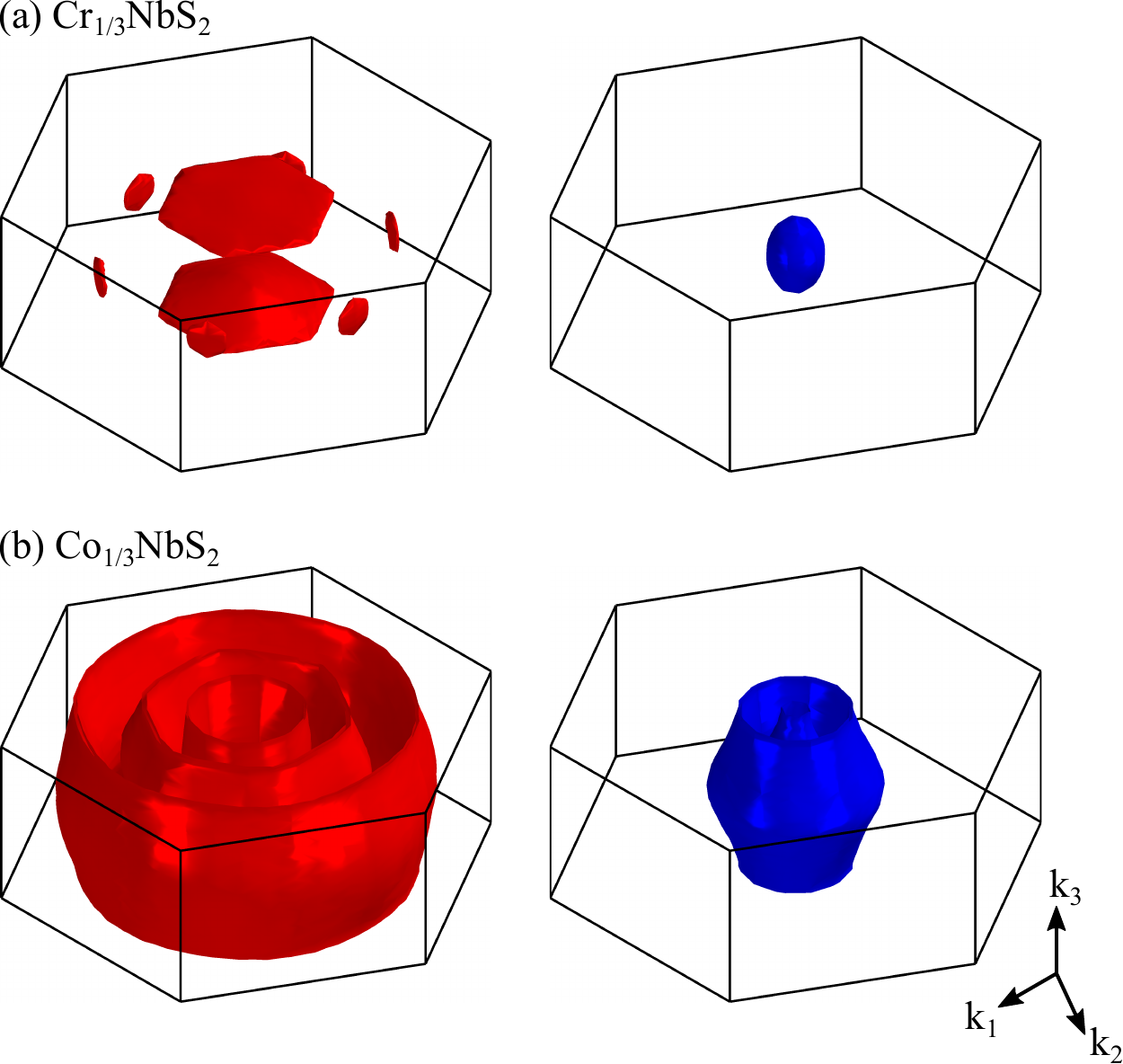}
  \caption{Fermi surfaces of (a) Cr$_{1/3}$NbS$_2$ and (b) Co$_{1/3}$NbS$_2$ showing contributions from the spin-up (red) and spin-down (blue) electrons.}
  \label{fig:tmdc_fs}
\end{figure}

It is notable that the systems with more than half-full 3$d$ shells ($N$~=~Fe, Co, and Ni) have an antiferromagnetically-ordered ground state, rather than the ferromagnetic one in our calculations. 
Insight into this is provided by the topology of the Fermi surfaces across the series, examples of which are shown in Fig.~\ref{fig:tmdc_fs}. 
(Further such Fermi surfaces can be seen in Ref.~\cite{sm}.)
For $N$~=~Fe, Co, and Ni the Fermi surface has a two-dimensional character.
This reduced dimensionality leads to an increased influence of electronic instabilities, caused for example by Fermi-surface nesting or strong Coulomb interactions of the 4$d$ band of the Nb atom~\cite{guller2016}.
This can cause (antiferromagnetic) spin-density-wave formation and a consequent reconstruction of the Fermi surface.
In contrast, the materials with $N$~=~Ti, V, and Cr appear far more three dimensional, leading to a ferromagnetic exchange interaction, with the $N$~=~Mn system lying somewhere in between.

Our first principles study into the magnetic properties of the intercalated TMDCs demonstrates that a rigid-band picture of the electronic structure largely explains the magnetism of the system.
Using a simple ferromagnetic model for all the materials in this series, we are able to accurately reproduce experimental observations and describe the underlying electronic structure.
We find that the degeneracy of the 3$d$ orbitals is broken such that the electron filling pattern, distinct from that of the isolated intercalant, is consistent with the local symmetry of the crystal and correctly predicts the magnetic moment.
The calculated Fermi surfaces provide insight into the exchange interaction that manifests in the systems.
Our work demonstrates that, in this class of materials, not only do first principles calculations lead to unique insights, but that they are sufficiently successful at predicting experimental results that they could be used as a cost-effective technique for screening materials for desired properties before crystal synthesis is required.

We thank  Amelia~Hall,  Chennan Wang and Li Yu for useful discussion.
We acknowledge the support of EPSRC (UK) [EP/N032128/1]  and  Durham Hamilton HPC. Research data will be made available via \textcolor{red}{XXX}. 

\bibliography{bib}

\begin{thebibliography}{49}%
\makeatletter
\providecommand \@ifxundefined [1]{%
 \@ifx{#1\undefined}
}%
\providecommand \@ifnum [1]{%
 \ifnum #1\expandafter \@firstoftwo
 \else \expandafter \@secondoftwo
 \fi
}%
\providecommand \@ifx [1]{%
 \ifx #1\expandafter \@firstoftwo
 \else \expandafter \@secondoftwo
 \fi
}%
\providecommand \natexlab [1]{#1}%
\providecommand \enquote  [1]{``#1''}%
\providecommand \bibnamefont  [1]{#1}%
\providecommand \bibfnamefont [1]{#1}%
\providecommand \citenamefont [1]{#1}%
\providecommand \href@noop [0]{\@secondoftwo}%
\providecommand \href [0]{\begingroup \@sanitize@url \@href}%
\providecommand \@href[1]{\@@startlink{#1}\@@href}%
\providecommand \@@href[1]{\endgroup#1\@@endlink}%
\providecommand \@sanitize@url [0]{\catcode `\\12\catcode `\$12\catcode
  `\&12\catcode `\#12\catcode `\^12\catcode `\_12\catcode `\%12\relax}%
\providecommand \@@startlink[1]{}%
\providecommand \@@endlink[0]{}%
\providecommand \url  [0]{\begingroup\@sanitize@url \@url }%
\providecommand \@url [1]{\endgroup\@href {#1}{\urlprefix }}%
\providecommand \urlprefix  [0]{URL }%
\providecommand \Eprint [0]{\href }%
\providecommand \doibase [0]{https://doi.org/}%
\providecommand \selectlanguage [0]{\@gobble}%
\providecommand \bibinfo  [0]{\@secondoftwo}%
\providecommand \bibfield  [0]{\@secondoftwo}%
\providecommand \translation [1]{[#1]}%
\providecommand \BibitemOpen [0]{}%
\providecommand \bibitemStop [0]{}%
\providecommand \bibitemNoStop [0]{.\EOS\space}%
\providecommand \EOS [0]{\spacefactor3000\relax}%
\providecommand \BibitemShut  [1]{\csname bibitem#1\endcsname}%
\let\auto@bib@innerbib\@empty
\bibitem [{\citenamefont {Manzeli}\ \emph {et~al.}(2017)\citenamefont
  {Manzeli}, \citenamefont {Ovchinnikov}, \citenamefont {Pasquier},
  \citenamefont {Yazyev},\ and\ \citenamefont {Kis}}]{manzeli20172d}%
  \BibitemOpen
  \bibfield  {author} {\bibinfo {author} {\bibfnamefont {S.}~\bibnamefont
  {Manzeli}}, \bibinfo {author} {\bibfnamefont {D.}~\bibnamefont
  {Ovchinnikov}}, \bibinfo {author} {\bibfnamefont {D.}~\bibnamefont
  {Pasquier}}, \bibinfo {author} {\bibfnamefont {O.~V.}\ \bibnamefont
  {Yazyev}},\ and\ \bibinfo {author} {\bibfnamefont {A.}~\bibnamefont {Kis}},\
  }\bibfield  {title} {\bibinfo {title} {2D transition metal dichalcogenides},\
  }\href {https://doi.org/10.1038/natrevmats.2017.33} {\bibfield  {journal}
  {\bibinfo  {journal} {Nat. Rev. Mater.}\ }\textbf {\bibinfo {volume} {2}},\
  \bibinfo {pages} {1} (\bibinfo {year} {2017})}\BibitemShut {NoStop}%
\bibitem [{\citenamefont {Wei}\ \emph {et~al.}(2018)\citenamefont {Wei},
  \citenamefont {Li}, \citenamefont {Xia}, \citenamefont {Cui}, \citenamefont
  {He}, \citenamefont {Xia},\ and\ \citenamefont {Li}}]{wei2018various}%
  \BibitemOpen
  \bibfield  {author} {\bibinfo {author} {\bibfnamefont {Z.}~\bibnamefont
  {Wei}}, \bibinfo {author} {\bibfnamefont {B.}~\bibnamefont {Li}}, \bibinfo
  {author} {\bibfnamefont {C.}~\bibnamefont {Xia}}, \bibinfo {author}
  {\bibfnamefont {Y.}~\bibnamefont {Cui}}, \bibinfo {author} {\bibfnamefont
  {J.}~\bibnamefont {He}}, \bibinfo {author} {\bibfnamefont {J.-B.}\
  \bibnamefont {Xia}},\ and\ \bibinfo {author} {\bibfnamefont {J.}~\bibnamefont
  {Li}},\ }\bibfield  {title} {\bibinfo {title} {Various structures of 2D
  transition-metal dichalcogenides and their applications},\ }\href
  {https://doi.org/10.1002/smtd.201800094} {\bibfield  {journal} {\bibinfo
  {journal} {Small Methods}\ }\textbf {\bibinfo {volume} {2}},\ \bibinfo
  {pages} {1800094} (\bibinfo {year} {2018})}\BibitemShut {NoStop}%
\bibitem [{\citenamefont {Wang}\ \emph {et~al.}(2012)\citenamefont {Wang},
  \citenamefont {Kalantar-Zadeh}, \citenamefont {Kis}, \citenamefont
  {Coleman},\ and\ \citenamefont {Strano}}]{wang2012electronics}%
  \BibitemOpen
  \bibfield  {author} {\bibinfo {author} {\bibfnamefont {Q.~H.}\ \bibnamefont
  {Wang}}, \bibinfo {author} {\bibfnamefont {K.}~\bibnamefont
  {Kalantar-Zadeh}}, \bibinfo {author} {\bibfnamefont {A.}~\bibnamefont {Kis}},
  \bibinfo {author} {\bibfnamefont {J.~N.}\ \bibnamefont {Coleman}},\ and\
  \bibinfo {author} {\bibfnamefont {M.~S.}\ \bibnamefont {Strano}},\ }\bibfield
   {title} {\bibinfo {title} {Electronics and optoelectronics of
  two-dimensional transition metal dichalcogenides},\ }\href
  {https://doi.org/10.1038/nnano.2012.193} {\bibfield  {journal} {\bibinfo
  {journal} {Nat. Nanotechnol.}\ }\textbf {\bibinfo {volume} {7}},\ \bibinfo
  {pages} {699} (\bibinfo {year} {2012})}\BibitemShut {NoStop}%
\bibitem [{\citenamefont {Mak}\ \emph {et~al.}(2010)\citenamefont {Mak},
  \citenamefont {Lee}, \citenamefont {Hone}, \citenamefont {Shan},\ and\
  \citenamefont {Heinz}}]{mak2010atomically}%
  \BibitemOpen
  \bibfield  {author} {\bibinfo {author} {\bibfnamefont {K.~F.}\ \bibnamefont
  {Mak}}, \bibinfo {author} {\bibfnamefont {C.}~\bibnamefont {Lee}}, \bibinfo
  {author} {\bibfnamefont {J.}~\bibnamefont {Hone}}, \bibinfo {author}
  {\bibfnamefont {J.}~\bibnamefont {Shan}},\ and\ \bibinfo {author}
  {\bibfnamefont {T.~F.}\ \bibnamefont {Heinz}},\ }\bibfield  {title} {\bibinfo
  {title} {Atomically thin MoS$_2$: a new direct-gap semiconductor},\ }\href
  {https://doi.org/10.1103/PhysRevLett.105.136805} {\bibfield  {journal}
  {\bibinfo  {journal} {Phys. Rev. Lett}\ }\textbf {\bibinfo {volume} {105}},\
  \bibinfo {pages} {136805} (\bibinfo {year} {2010})}\BibitemShut {NoStop}%
\bibitem [{\citenamefont {Wilson}\ and\ \citenamefont
  {Yoffe}(1969)}]{wilson1969transition}%
  \BibitemOpen
  \bibfield  {author} {\bibinfo {author} {\bibfnamefont {J.~A.}\ \bibnamefont
  {Wilson}}\ and\ \bibinfo {author} {\bibfnamefont {A.}~\bibnamefont {Yoffe}},\
  }\bibfield  {title} {\bibinfo {title} {The transition metal dichalcogenides
  discussion and interpretation of the observed optical, electrical and
  structural properties},\ }\href@noop {} {\bibfield  {journal} {\bibinfo
  {journal} {Advances in Physics}\ }\textbf {\bibinfo {volume} {18}},\ \bibinfo
  {pages} {193} (\bibinfo {year} {1969})}\BibitemShut {NoStop}%
\bibitem [{\citenamefont {Moncton}\ \emph {et~al.}(1975)\citenamefont
  {Moncton}, \citenamefont {Axe},\ and\ \citenamefont
  {DiSalvo}}]{moncton1975study}%
  \BibitemOpen
  \bibfield  {author} {\bibinfo {author} {\bibfnamefont {D.}~\bibnamefont
  {Moncton}}, \bibinfo {author} {\bibfnamefont {J.}~\bibnamefont {Axe}},\ and\
  \bibinfo {author} {\bibfnamefont {F.}~\bibnamefont {DiSalvo}},\ }\bibfield
  {title} {\bibinfo {title} {Study of superlattice formation in 2H-NbSe$_2$ and
  2H-TaSe$_2$ by neutron scattering},\ }\href
  {https://doi.org/10.1103/PhysRevLett.34.734} {\bibfield  {journal} {\bibinfo
  {journal} {Phys. Rev. Lett.}\ }\textbf {\bibinfo {volume} {34}},\ \bibinfo
  {pages} {734} (\bibinfo {year} {1975})}\BibitemShut {NoStop}%
\bibitem [{\citenamefont {Rajora}\ and\ \citenamefont
  {Curzon}(1987)}]{rajora1987preparation}%
  \BibitemOpen
  \bibfield  {author} {\bibinfo {author} {\bibfnamefont {O.}~\bibnamefont
  {Rajora}}\ and\ \bibinfo {author} {\bibfnamefont {A.}~\bibnamefont
  {Curzon}},\ }\bibfield  {title} {\bibinfo {title} {The preparation and X-ray
  diffraction study of the layer materials NbS$_x$Se$_{2-x}$ for 0$\leq$x$\leq$
  2},\ }\href {https://doi.org/10.1002/pssa.2210990108} {\bibfield  {journal}
  {\bibinfo  {journal} {physica status solidi (a)}\ }\textbf {\bibinfo {volume}
  {99}},\ \bibinfo {pages} {65} (\bibinfo {year} {1987})}\BibitemShut {NoStop}%
\bibitem [{\citenamefont {Naito}\ and\ \citenamefont
  {Tanaka}(1982)}]{naito1982electrical}%
  \BibitemOpen
  \bibfield  {author} {\bibinfo {author} {\bibfnamefont {M.}~\bibnamefont
  {Naito}}\ and\ \bibinfo {author} {\bibfnamefont {S.}~\bibnamefont {Tanaka}},\
  }\bibfield  {title} {\bibinfo {title} {Electrical transport properties in
  2H-NbS$_2$,-NbSe$_2$,-TaS$_2$ and-TaSe$_2$},\ }\href
  {https://doi.org/10.1143/JPSJ.51.219} {\bibfield  {journal} {\bibinfo
  {journal} {J. Phys. Soc. Jpn.}\ }\textbf {\bibinfo {volume} {51}},\ \bibinfo
  {pages} {219} (\bibinfo {year} {1982})}\BibitemShut {NoStop}%
\bibitem [{\citenamefont {Tissen}\ \emph {et~al.}(2013)\citenamefont {Tissen},
  \citenamefont {Osorio}, \citenamefont {Brison}, \citenamefont {Nemes},
  \citenamefont {Garc{\'\i}a-Hern{\'a}ndez}, \citenamefont {Cario},
  \citenamefont {Rodiere}, \citenamefont {Vieira},\ and\ \citenamefont
  {Suderow}}]{tissen2013pressure}%
  \BibitemOpen
  \bibfield  {author} {\bibinfo {author} {\bibfnamefont {V.}~\bibnamefont
  {Tissen}}, \bibinfo {author} {\bibfnamefont {M.}~\bibnamefont {Osorio}},
  \bibinfo {author} {\bibfnamefont {J.-P.}\ \bibnamefont {Brison}}, \bibinfo
  {author} {\bibfnamefont {N.}~\bibnamefont {Nemes}}, \bibinfo {author}
  {\bibfnamefont {M.}~\bibnamefont {Garc{\'\i}a-Hern{\'a}ndez}}, \bibinfo
  {author} {\bibfnamefont {L.}~\bibnamefont {Cario}}, \bibinfo {author}
  {\bibfnamefont {P.}~\bibnamefont {Rodiere}}, \bibinfo {author} {\bibfnamefont
  {S.}~\bibnamefont {Vieira}},\ and\ \bibinfo {author} {\bibfnamefont
  {H.}~\bibnamefont {Suderow}},\ }\bibfield  {title} {\bibinfo {title}
  {Pressure dependence of superconducting critical temperature and upper
  critical field of 2$H$-NbS$_2$},\ }\href
  {https://doi.org/10.1103/PhysRevB.87.134502} {\bibfield  {journal} {\bibinfo
  {journal} {Phys. Rev. B.}\ }\textbf {\bibinfo {volume} {87}},\ \bibinfo
  {pages} {134502} (\bibinfo {year} {2013})}\BibitemShut {NoStop}%
\bibitem [{\citenamefont {Heil}\ \emph {et~al.}(2017)\citenamefont {Heil},
  \citenamefont {Ponc{\'e}}, \citenamefont {Lambert}, \citenamefont {Schlipf},
  \citenamefont {Margine},\ and\ \citenamefont {Giustino}}]{heil2017origin}%
  \BibitemOpen
  \bibfield  {author} {\bibinfo {author} {\bibfnamefont {C.}~\bibnamefont
  {Heil}}, \bibinfo {author} {\bibfnamefont {S.}~\bibnamefont {Ponc{\'e}}},
  \bibinfo {author} {\bibfnamefont {H.}~\bibnamefont {Lambert}}, \bibinfo
  {author} {\bibfnamefont {M.}~\bibnamefont {Schlipf}}, \bibinfo {author}
  {\bibfnamefont {E.~R.}\ \bibnamefont {Margine}},\ and\ \bibinfo {author}
  {\bibfnamefont {F.}~\bibnamefont {Giustino}},\ }\bibfield  {title} {\bibinfo
  {title} {Origin of superconductivity and latent charge density wave in
  NbS$_2$},\ }\href {https://doi.org/10.1103/PhysRevLett.119.087003} {\bibfield
   {journal} {\bibinfo  {journal} {Phys. Rev. Lett}\ }\textbf {\bibinfo
  {volume} {119}},\ \bibinfo {pages} {087003} (\bibinfo {year}
  {2017})}\BibitemShut {NoStop}%
\bibitem [{\citenamefont {Witteveen}\ \emph {et~al.}(2021)\citenamefont
  {Witteveen}, \citenamefont {G{\'o}rnicka}, \citenamefont {Chang},
  \citenamefont {M{\aa}nsson}, \citenamefont {Klimczuk},\ and\ \citenamefont
  {von Rohr}}]{witteveen2021polytypism}%
  \BibitemOpen
  \bibfield  {author} {\bibinfo {author} {\bibfnamefont {C.}~\bibnamefont
  {Witteveen}}, \bibinfo {author} {\bibfnamefont {K.}~\bibnamefont
  {G{\'o}rnicka}}, \bibinfo {author} {\bibfnamefont {J.}~\bibnamefont {Chang}},
  \bibinfo {author} {\bibfnamefont {M.}~\bibnamefont {M{\aa}nsson}}, \bibinfo
  {author} {\bibfnamefont {T.}~\bibnamefont {Klimczuk}},\ and\ \bibinfo
  {author} {\bibfnamefont {F.~O.}\ \bibnamefont {von Rohr}},\ }\bibfield
  {title} {\bibinfo {title} {Polytypism and superconductivity in the NbS$_2$
  system},\ }\href {https://doi.org/10.1039/d0dt03636f} {\bibfield  {journal}
  {\bibinfo  {journal} {Dalton Trans.}\ }\textbf {\bibinfo {volume} {50}},\
  \bibinfo {pages} {3216} (\bibinfo {year} {2021})}\BibitemShut {NoStop}%
\bibitem [{\citenamefont {Castellanos-Gomez}\ \emph {et~al.}(2012)\citenamefont
  {Castellanos-Gomez}, \citenamefont {Barkelid}, \citenamefont {Goossens},
  \citenamefont {Calado}, \citenamefont {van~der Zant},\ and\ \citenamefont
  {Steele}}]{castellanos2012laser}%
  \BibitemOpen
  \bibfield  {author} {\bibinfo {author} {\bibfnamefont {A.}~\bibnamefont
  {Castellanos-Gomez}}, \bibinfo {author} {\bibfnamefont {M.}~\bibnamefont
  {Barkelid}}, \bibinfo {author} {\bibfnamefont {A.}~\bibnamefont {Goossens}},
  \bibinfo {author} {\bibfnamefont {V.~E.}\ \bibnamefont {Calado}}, \bibinfo
  {author} {\bibfnamefont {H.~S.}\ \bibnamefont {van~der Zant}},\ and\ \bibinfo
  {author} {\bibfnamefont {G.~A.}\ \bibnamefont {Steele}},\ }\bibfield  {title}
  {\bibinfo {title} {Laser-thinning of MoS$_2$: on demand generation of a
  single-layer semiconductor},\ }\href {https://doi.org/10.1021/nl301164v}
  {\bibfield  {journal} {\bibinfo  {journal} {Nano Letters}\ }\textbf {\bibinfo
  {volume} {12}},\ \bibinfo {pages} {3187} (\bibinfo {year}
  {2012})}\BibitemShut {NoStop}%
\bibitem [{\citenamefont {Ganatra}\ and\ \citenamefont
  {Zhang}(2014)}]{ganatra2014few}%
  \BibitemOpen
  \bibfield  {author} {\bibinfo {author} {\bibfnamefont {R.}~\bibnamefont
  {Ganatra}}\ and\ \bibinfo {author} {\bibfnamefont {Q.}~\bibnamefont
  {Zhang}},\ }\bibfield  {title} {\bibinfo {title} {Few-layer MoS$_2$: a
  promising layered semiconductor},\ }\href {https://doi.org/10.1021/nn405938z}
  {\bibfield  {journal} {\bibinfo  {journal} {ACS Nano}\ }\textbf {\bibinfo
  {volume} {8}},\ \bibinfo {pages} {4074} (\bibinfo {year} {2014})}\BibitemShut
  {NoStop}%
\bibitem [{\citenamefont {Ramasubramaniam}\ \emph {et~al.}(2011)\citenamefont
  {Ramasubramaniam}, \citenamefont {Naveh},\ and\ \citenamefont
  {Towe}}]{ramasubramaniam2011tunable}%
  \BibitemOpen
  \bibfield  {author} {\bibinfo {author} {\bibfnamefont {A.}~\bibnamefont
  {Ramasubramaniam}}, \bibinfo {author} {\bibfnamefont {D.}~\bibnamefont
  {Naveh}},\ and\ \bibinfo {author} {\bibfnamefont {E.}~\bibnamefont {Towe}},\
  }\bibfield  {title} {\bibinfo {title} {Tunable band gaps in bilayer
  transition-metal dichalcogenides},\ }\href
  {https://doi.org/10.1103/PhysRevB.84.205325} {\bibfield  {journal} {\bibinfo
  {journal} {Physical Review B}\ }\textbf {\bibinfo {volume} {84}},\ \bibinfo
  {pages} {205325} (\bibinfo {year} {2011})}\BibitemShut {NoStop}%
\bibitem [{\citenamefont {Battaglia}\ \emph {et~al.}(2007)\citenamefont
  {Battaglia}, \citenamefont {Cercellier}, \citenamefont {Despont},
  \citenamefont {Monney}, \citenamefont {Prester}, \citenamefont {Berger},
  \citenamefont {Forr{\'o}}, \citenamefont {Garnier},\ and\ \citenamefont
  {Aebi}}]{battaglia2007non}%
  \BibitemOpen
  \bibfield  {author} {\bibinfo {author} {\bibfnamefont {C.}~\bibnamefont
  {Battaglia}}, \bibinfo {author} {\bibfnamefont {H.}~\bibnamefont
  {Cercellier}}, \bibinfo {author} {\bibfnamefont {L.}~\bibnamefont {Despont}},
  \bibinfo {author} {\bibfnamefont {C.}~\bibnamefont {Monney}}, \bibinfo
  {author} {\bibfnamefont {M.}~\bibnamefont {Prester}}, \bibinfo {author}
  {\bibfnamefont {H.}~\bibnamefont {Berger}}, \bibinfo {author} {\bibfnamefont
  {L.}~\bibnamefont {Forr{\'o}}}, \bibinfo {author} {\bibfnamefont
  {M.}~\bibnamefont {Garnier}},\ and\ \bibinfo {author} {\bibfnamefont
  {P.}~\bibnamefont {Aebi}},\ }\bibfield  {title} {\bibinfo {title}
  {Non-uniform doping across the Fermi surface of NbS$_2$ intercalates},\
  }\href {https://doi.org/10.1140/epjb/e2007-00188-1} {\bibfield  {journal}
  {\bibinfo  {journal} {The European Physical Journal B}\ }\textbf {\bibinfo
  {volume} {57}},\ \bibinfo {pages} {385} (\bibinfo {year} {2007})}\BibitemShut
  {NoStop}%
\bibitem [{\citenamefont {Clark}(1976)}]{clark1976structural}%
  \BibitemOpen
  \bibfield  {author} {\bibinfo {author} {\bibfnamefont {W.}~\bibnamefont
  {Clark}},\ }\bibfield  {title} {\bibinfo {title} {Structural and
  photoemission studies of some transition metal intercalates of NbS$_2$},\
  }\href {https://doi.org/10.1088/0022-3719/9/24/005} {\bibfield  {journal}
  {\bibinfo  {journal} {Journal of Physics C: Solid State Physics}\ }\textbf
  {\bibinfo {volume} {9}},\ \bibinfo {pages} {L693} (\bibinfo {year}
  {1976})}\BibitemShut {NoStop}%
\bibitem [{\citenamefont {Hicken}\ \emph
  {et~al.}(2022{\natexlab{a}})\citenamefont {Hicken}, \citenamefont {Hawkhead},
  \citenamefont {Wilson}, \citenamefont {Huddart}, \citenamefont {Hall},
  \citenamefont {Balakrishnan}, \citenamefont {Wang}, \citenamefont {Pratt},
  \citenamefont {Clark},\ and\ \citenamefont {Lancaster}}]{hicken2022energy}%
  \BibitemOpen
  \bibfield  {author} {\bibinfo {author} {\bibfnamefont {T.}~\bibnamefont
  {Hicken}}, \bibinfo {author} {\bibfnamefont {Z.}~\bibnamefont {Hawkhead}},
  \bibinfo {author} {\bibfnamefont {M.}~\bibnamefont {Wilson}}, \bibinfo
  {author} {\bibfnamefont {B.}~\bibnamefont {Huddart}}, \bibinfo {author}
  {\bibfnamefont {A.}~\bibnamefont {Hall}}, \bibinfo {author} {\bibfnamefont
  {G.}~\bibnamefont {Balakrishnan}}, \bibinfo {author} {\bibfnamefont
  {C.}~\bibnamefont {Wang}}, \bibinfo {author} {\bibfnamefont {F.}~\bibnamefont
  {Pratt}}, \bibinfo {author} {\bibfnamefont {S.}~\bibnamefont {Clark}},\ and\
  \bibinfo {author} {\bibfnamefont {T.}~\bibnamefont {Lancaster}},\ }\bibfield
  {title} {\bibinfo {title} {Energy-gap driven low-temperature magnetic and
  transport properties in Cr$_{1/3}$$M$S$_2$ ($M$= Nb, Ta)},\ }\href
  {https://doi.org/10.1103/PhysRevB.105.L060407} {\bibfield  {journal}
  {\bibinfo  {journal} {Phys. Rev. B.}\ }\textbf {\bibinfo {volume} {105}},\
  \bibinfo {pages} {L060407} (\bibinfo {year}
  {2022}{\natexlab{a}})}\BibitemShut {NoStop}%
\bibitem [{\citenamefont {Ghimire}\ \emph {et~al.}(2013)\citenamefont
  {Ghimire}, \citenamefont {McGuire}, \citenamefont {Parker}, \citenamefont
  {Sipos}, \citenamefont {Tang}, \citenamefont {Yan}, \citenamefont {Sales},\
  and\ \citenamefont {Mandrus}}]{ghimire2013magnetic}%
  \BibitemOpen
  \bibfield  {author} {\bibinfo {author} {\bibfnamefont {N.}~\bibnamefont
  {Ghimire}}, \bibinfo {author} {\bibfnamefont {M.~A.}\ \bibnamefont
  {McGuire}}, \bibinfo {author} {\bibfnamefont {D.~S.}\ \bibnamefont {Parker}},
  \bibinfo {author} {\bibfnamefont {B.}~\bibnamefont {Sipos}}, \bibinfo
  {author} {\bibfnamefont {S.}~\bibnamefont {Tang}}, \bibinfo {author}
  {\bibfnamefont {J.-Q.}\ \bibnamefont {Yan}}, \bibinfo {author} {\bibfnamefont
  {B.~C.}\ \bibnamefont {Sales}},\ and\ \bibinfo {author} {\bibfnamefont
  {D.}~\bibnamefont {Mandrus}},\ }\bibfield  {title} {\bibinfo {title}
  {Magnetic phase transition in single crystals of the chiral helimagnet
  Cr$_{1/3}$NbS$_2$},\ }\href {https://doi.org/10.1103/PhysRevB.87.104403}
  {\bibfield  {journal} {\bibinfo  {journal} {Phys. Rev. B.}\ }\textbf
  {\bibinfo {volume} {87}},\ \bibinfo {pages} {104403} (\bibinfo {year}
  {2013})}\BibitemShut {NoStop}%
\bibitem [{\citenamefont {Parkin}\ and\ \citenamefont
  {Friend}(1980{\natexlab{a}})}]{parkin19803i}%
  \BibitemOpen
  \bibfield  {author} {\bibinfo {author} {\bibfnamefont {S.}~\bibnamefont
  {Parkin}}\ and\ \bibinfo {author} {\bibfnamefont {R.}~\bibnamefont
  {Friend}},\ }\bibfield  {title} {\bibinfo {title} {3$d$ transition-metal
  intercalates of the niobium and tantalum dichalcogenides. I. Magnetic
  properties},\ }\href {https://doi.org/10.1080/13642818008245370} {\bibfield
  {journal} {\bibinfo  {journal} {Philosophical Magazine B}\ }\textbf {\bibinfo
  {volume} {41}},\ \bibinfo {pages} {65} (\bibinfo {year}
  {1980}{\natexlab{a}})}\BibitemShut {NoStop}%
\bibitem [{\citenamefont {Parkin}\ and\ \citenamefont
  {Friend}(1980{\natexlab{b}})}]{parkin19803ii}%
  \BibitemOpen
  \bibfield  {author} {\bibinfo {author} {\bibfnamefont {S.}~\bibnamefont
  {Parkin}}\ and\ \bibinfo {author} {\bibfnamefont {R.}~\bibnamefont
  {Friend}},\ }\bibfield  {title} {\bibinfo {title} {3$d$ transition-metal
  intercalates of the niobium and tantalum dichalcogenides. II. Transport
  properties},\ }\href {https://doi.org/10.1080/13642818008245371} {\bibfield
  {journal} {\bibinfo  {journal} {Philosophical Magazine B}\ }\textbf {\bibinfo
  {volume} {41}},\ \bibinfo {pages} {95} (\bibinfo {year}
  {1980}{\natexlab{b}})}\BibitemShut {NoStop}%
\bibitem [{\citenamefont {Hall}\ \emph {et~al.}(2021)\citenamefont {Hall},
  \citenamefont {Khalyavin}, \citenamefont {Manuel}, \citenamefont {Mayoh},
  \citenamefont {Orlandi}, \citenamefont {Petrenko}, \citenamefont {Lees},\
  and\ \citenamefont {Balakrishnan}}]{hall2021magnetic}%
  \BibitemOpen
  \bibfield  {author} {\bibinfo {author} {\bibfnamefont {A.}~\bibnamefont
  {Hall}}, \bibinfo {author} {\bibfnamefont {D.}~\bibnamefont {Khalyavin}},
  \bibinfo {author} {\bibfnamefont {P.}~\bibnamefont {Manuel}}, \bibinfo
  {author} {\bibfnamefont {D.}~\bibnamefont {Mayoh}}, \bibinfo {author}
  {\bibfnamefont {F.}~\bibnamefont {Orlandi}}, \bibinfo {author} {\bibfnamefont
  {O.}~\bibnamefont {Petrenko}}, \bibinfo {author} {\bibfnamefont
  {M.}~\bibnamefont {Lees}},\ and\ \bibinfo {author} {\bibfnamefont
  {G.}~\bibnamefont {Balakrishnan}},\ }\bibfield  {title} {\bibinfo {title}
  {Magnetic structure investigation of the intercalated transition metal
  dichalcogenide V$_{1/3}$NbS$_2$},\ }\href
  {https://doi.org/10.1103/PhysRevB.103.174431} {\bibfield  {journal} {\bibinfo
   {journal} {Phys. Rev. B.}\ }\textbf {\bibinfo {volume} {103}},\ \bibinfo
  {pages} {174431} (\bibinfo {year} {2021})}\BibitemShut {NoStop}%
\bibitem [{\citenamefont {Moriya}\ and\ \citenamefont
  {Miyadai}(1982)}]{moriya1982evidence}%
  \BibitemOpen
  \bibfield  {author} {\bibinfo {author} {\bibfnamefont {T.}~\bibnamefont
  {Moriya}}\ and\ \bibinfo {author} {\bibfnamefont {T.}~\bibnamefont
  {Miyadai}},\ }\bibfield  {title} {\bibinfo {title} {Evidence for the helical
  spin structure due to antisymmetric exchange interaction in
  Cr$_{1/3}$NbS$_2$},\ }\href {https://doi.org/10.1016/0038-1098(82)91006-7}
  {\bibfield  {journal} {\bibinfo  {journal} {Solid State Communications}\
  }\textbf {\bibinfo {volume} {42}},\ \bibinfo {pages} {209} (\bibinfo {year}
  {1982})}\BibitemShut {NoStop}%
\bibitem [{\citenamefont {Kousaka}\ \emph {et~al.}(2009)\citenamefont
  {Kousaka}, \citenamefont {Nakao}, \citenamefont {Furukawa},\ and\
  \citenamefont {Akimitsu}}]{kousaka2009chiral}%
  \BibitemOpen
  \bibfield  {author} {\bibinfo {author} {\bibfnamefont {Y.}~\bibnamefont
  {Kousaka}}, \bibinfo {author} {\bibfnamefont {Y.}~\bibnamefont {Nakao}},
  \bibinfo {author} {\bibfnamefont {H.}~\bibnamefont {Furukawa}},\ and\
  \bibinfo {author} {\bibfnamefont {J.}~\bibnamefont {Akimitsu}},\ }\bibfield
  {title} {\bibinfo {title} {Chiral Helimagnetic Order in $T_{1/3}$NbS$_2$
  ($T$= Cr, Mn)},\ }\href {https://doi.org/10.1016/j.nima.2008.11.040}
  {\bibfield  {journal} {\bibinfo  {journal} {Activity Report on Neutron
  Scattering Research: Experimental Reports}\ }\textbf {\bibinfo {volume}
  {16}},\ \bibinfo {pages} {814} (\bibinfo {year} {2009})}\BibitemShut
  {NoStop}%
\bibitem [{\citenamefont {Hall}\ \emph {et~al.}(2022)\citenamefont {Hall},
  \citenamefont {Loudon}, \citenamefont {Midgley}, \citenamefont
  {Twitchett-Harrison}, \citenamefont {Holt}, \citenamefont {Mayoh},
  \citenamefont {Tidey}, \citenamefont {Han}, \citenamefont {Lees},\ and\
  \citenamefont {Balakrishnan}}]{hall2022comparative}%
  \BibitemOpen
  \bibfield  {author} {\bibinfo {author} {\bibfnamefont {A.}~\bibnamefont
  {Hall}}, \bibinfo {author} {\bibfnamefont {J.}~\bibnamefont {Loudon}},
  \bibinfo {author} {\bibfnamefont {P.}~\bibnamefont {Midgley}}, \bibinfo
  {author} {\bibfnamefont {A.}~\bibnamefont {Twitchett-Harrison}}, \bibinfo
  {author} {\bibfnamefont {S.}~\bibnamefont {Holt}}, \bibinfo {author}
  {\bibfnamefont {D.}~\bibnamefont {Mayoh}}, \bibinfo {author} {\bibfnamefont
  {J.}~\bibnamefont {Tidey}}, \bibinfo {author} {\bibfnamefont
  {Y.}~\bibnamefont {Han}}, \bibinfo {author} {\bibfnamefont {M.}~\bibnamefont
  {Lees}},\ and\ \bibinfo {author} {\bibfnamefont {G.}~\bibnamefont
  {Balakrishnan}},\ }\bibfield  {title} {\bibinfo {title} {Comparative study of
  the structural and magnetic properties of Mn$_{1/3}$NbS$_2$ and
  Cr$_{1/3}$NbS$_2$},\ }\href
  {https://doi.org/10.1103/PhysRevMaterials.6.024407} {\bibfield  {journal}
  {\bibinfo  {journal} {Phys. Rev. Materials}\ }\textbf {\bibinfo {volume}
  {6}},\ \bibinfo {pages} {024407} (\bibinfo {year} {2022})}\BibitemShut
  {NoStop}%
\bibitem [{\citenamefont {Van~Laar}\ \emph {et~al.}(1971)\citenamefont
  {Van~Laar}, \citenamefont {Rietveld},\ and\ \citenamefont
  {Ijdo}}]{van1971magnetic}%
  \BibitemOpen
  \bibfield  {author} {\bibinfo {author} {\bibfnamefont {B.}~\bibnamefont
  {Van~Laar}}, \bibinfo {author} {\bibfnamefont {H.}~\bibnamefont {Rietveld}},\
  and\ \bibinfo {author} {\bibfnamefont {D.}~\bibnamefont {Ijdo}},\ }\bibfield
  {title} {\bibinfo {title} {Magnetic and crystallographic structures of
  Me$_x$NbS$_2$ and Me$_x$TaS$_2$},\ }\href
  {https://doi.org/10.1016/0022-4596(71)90019-3} {\bibfield  {journal}
  {\bibinfo  {journal} {Journal of Solid State Chemistry}\ }\textbf {\bibinfo
  {volume} {3}},\ \bibinfo {pages} {154} (\bibinfo {year} {1971})}\BibitemShut
  {NoStop}%
\bibitem [{\citenamefont {Little}\ \emph {et~al.}(2020)\citenamefont {Little},
  \citenamefont {Lee}, \citenamefont {John}, \citenamefont {Doyle},
  \citenamefont {Maniv}, \citenamefont {Nair}, \citenamefont {Chen},
  \citenamefont {Rees}, \citenamefont {Venderbos}, \citenamefont {Fernandes}
  \emph {et~al.}}]{little2020three}%
  \BibitemOpen
  \bibfield  {author} {\bibinfo {author} {\bibfnamefont {A.}~\bibnamefont
  {Little}}, \bibinfo {author} {\bibfnamefont {C.}~\bibnamefont {Lee}},
  \bibinfo {author} {\bibfnamefont {C.}~\bibnamefont {John}}, \bibinfo {author}
  {\bibfnamefont {S.}~\bibnamefont {Doyle}}, \bibinfo {author} {\bibfnamefont
  {E.}~\bibnamefont {Maniv}}, \bibinfo {author} {\bibfnamefont {N.~L.}\
  \bibnamefont {Nair}}, \bibinfo {author} {\bibfnamefont {W.}~\bibnamefont
  {Chen}}, \bibinfo {author} {\bibfnamefont {D.}~\bibnamefont {Rees}}, \bibinfo
  {author} {\bibfnamefont {J.~W.}\ \bibnamefont {Venderbos}}, \bibinfo {author}
  {\bibfnamefont {R.~M.}\ \bibnamefont {Fernandes}}, \emph {et~al.},\
  }\bibfield  {title} {\bibinfo {title} {Three-state nematicity in the
  triangular lattice antiferromagnet Fe$_{1/3}$NbS$_2$},\ }\href
  {https://doi.org/10.1103/PhysRevResearch.2.043020} {\bibfield  {journal}
  {\bibinfo  {journal} {Nature Materials}\ }\textbf {\bibinfo {volume} {19}},\
  \bibinfo {pages} {1062} (\bibinfo {year} {2020})}\BibitemShut {NoStop}%
\bibitem [{\citenamefont {Xie}\ \emph {et~al.}(2022)\citenamefont {Xie},
  \citenamefont {Husremović}, \citenamefont {Gonzalez}, \citenamefont
  {Craig},\ and\ \citenamefont {Bediako}}]{xie2022structure}%
  \BibitemOpen
  \bibfield  {author} {\bibinfo {author} {\bibfnamefont {L.~S.}\ \bibnamefont
  {Xie}}, \bibinfo {author} {\bibfnamefont {S.}~\bibnamefont {Husremović}},
  \bibinfo {author} {\bibfnamefont {O.}~\bibnamefont {Gonzalez}}, \bibinfo
  {author} {\bibfnamefont {I.~M.}\ \bibnamefont {Craig}},\ and\ \bibinfo
  {author} {\bibfnamefont {D.~K.}\ \bibnamefont {Bediako}},\ }\bibfield
  {title} {\bibinfo {title} {Structure and Magnetism of Iron- and
  Chromium-Intercalated Niobium and Tantalum Disulfides},\ }\href
  {https://doi.org/10.1021/jacs.1c12975} {\bibfield  {journal} {\bibinfo
  {journal} {Journal of the American Chemical Society}\ }\textbf {\bibinfo
  {volume} {144}},\ \bibinfo {pages} {9525} (\bibinfo {year}
  {2022})}\BibitemShut {NoStop}%
\bibitem [{\citenamefont {Parkin}\ \emph {et~al.}(1983)\citenamefont {Parkin},
  \citenamefont {Marseglia},\ and\ \citenamefont {Brown}}]{parkin1983magnetic}%
  \BibitemOpen
  \bibfield  {author} {\bibinfo {author} {\bibfnamefont {S.}~\bibnamefont
  {Parkin}}, \bibinfo {author} {\bibfnamefont {E.}~\bibnamefont {Marseglia}},\
  and\ \bibinfo {author} {\bibfnamefont {P.}~\bibnamefont {Brown}},\ }\bibfield
   {title} {\bibinfo {title} {Magnetic structure of Co$_{1/3}$NbS$_2$ and
  Co$_{1/3}$TaS$_2$},\ }\href {https://doi.org/10.1088/0022-3719/16/14/016}
  {\bibfield  {journal} {\bibinfo  {journal} {Journal of Physics C: Solid State
  Physics}\ }\textbf {\bibinfo {volume} {16}},\ \bibinfo {pages} {2765}
  (\bibinfo {year} {1983})}\BibitemShut {NoStop}%
\bibitem [{\citenamefont {Tenasini}\ \emph {et~al.}(2020)\citenamefont
  {Tenasini}, \citenamefont {Martino}, \citenamefont {Ubrig}, \citenamefont
  {Ghimire}, \citenamefont {Berger}, \citenamefont {Zaharko}, \citenamefont
  {Wu}, \citenamefont {Mitchell}, \citenamefont {Martin}, \citenamefont
  {Forr{\'o}} \emph {et~al.}}]{tenasini2020giant}%
  \BibitemOpen
  \bibfield  {author} {\bibinfo {author} {\bibfnamefont {G.}~\bibnamefont
  {Tenasini}}, \bibinfo {author} {\bibfnamefont {E.}~\bibnamefont {Martino}},
  \bibinfo {author} {\bibfnamefont {N.}~\bibnamefont {Ubrig}}, \bibinfo
  {author} {\bibfnamefont {N.~J.}\ \bibnamefont {Ghimire}}, \bibinfo {author}
  {\bibfnamefont {H.}~\bibnamefont {Berger}}, \bibinfo {author} {\bibfnamefont
  {O.}~\bibnamefont {Zaharko}}, \bibinfo {author} {\bibfnamefont
  {F.}~\bibnamefont {Wu}}, \bibinfo {author} {\bibfnamefont {J.}~\bibnamefont
  {Mitchell}}, \bibinfo {author} {\bibfnamefont {I.}~\bibnamefont {Martin}},
  \bibinfo {author} {\bibfnamefont {L.}~\bibnamefont {Forr{\'o}}}, \emph
  {et~al.},\ }\bibfield  {title} {\bibinfo {title} {Giant anomalous Hall effect
  in quasi-two-dimensional layered antiferromagnet Co$_{1/3}$NbS$_2$},\ }\href
  {https://doi.org/10.1103/PhysRevResearch.2.023051} {\bibfield  {journal}
  {\bibinfo  {journal} {Physical Review Research}\ }\textbf {\bibinfo {volume}
  {2}},\ \bibinfo {pages} {023051} (\bibinfo {year} {2020})}\BibitemShut
  {NoStop}%
\bibitem [{\citenamefont {Miyadai}\ \emph {et~al.}(1983)\citenamefont
  {Miyadai}, \citenamefont {Kikuchi}, \citenamefont {Kondo}, \citenamefont
  {Sakka}, \citenamefont {Arai},\ and\ \citenamefont
  {Ishikawa}}]{miyadai1983magnetic}%
  \BibitemOpen
  \bibfield  {author} {\bibinfo {author} {\bibfnamefont {T.}~\bibnamefont
  {Miyadai}}, \bibinfo {author} {\bibfnamefont {K.}~\bibnamefont {Kikuchi}},
  \bibinfo {author} {\bibfnamefont {H.}~\bibnamefont {Kondo}}, \bibinfo
  {author} {\bibfnamefont {S.}~\bibnamefont {Sakka}}, \bibinfo {author}
  {\bibfnamefont {M.}~\bibnamefont {Arai}},\ and\ \bibinfo {author}
  {\bibfnamefont {Y.}~\bibnamefont {Ishikawa}},\ }\bibfield  {title} {\bibinfo
  {title} {Magnetic properties of Cr$_{1/3}$NbS$_2$},\ }\href
  {https://doi.org/https://doi.org/10.1143/JPSJ.52.1394} {\bibfield  {journal}
  {\bibinfo  {journal} {Journal of the Physical Society of Japan}\ }\textbf
  {\bibinfo {volume} {52}},\ \bibinfo {pages} {1394} (\bibinfo {year}
  {1983})}\BibitemShut {NoStop}%
\bibitem [{\citenamefont {Kousaka}\ \emph {et~al.}(2016)\citenamefont
  {Kousaka}, \citenamefont {Ogura}, \citenamefont {Zhang}, \citenamefont
  {Miao}, \citenamefont {Lee}, \citenamefont {Torii}, \citenamefont {Kamiyama},
  \citenamefont {Campo}, \citenamefont {Inoue},\ and\ \citenamefont
  {Akimitsu}}]{kousaka2016long}%
  \BibitemOpen
  \bibfield  {author} {\bibinfo {author} {\bibfnamefont {Y.}~\bibnamefont
  {Kousaka}}, \bibinfo {author} {\bibfnamefont {T.}~\bibnamefont {Ogura}},
  \bibinfo {author} {\bibfnamefont {J.}~\bibnamefont {Zhang}}, \bibinfo
  {author} {\bibfnamefont {P.}~\bibnamefont {Miao}}, \bibinfo {author}
  {\bibfnamefont {S.}~\bibnamefont {Lee}}, \bibinfo {author} {\bibfnamefont
  {S.}~\bibnamefont {Torii}}, \bibinfo {author} {\bibfnamefont
  {T.}~\bibnamefont {Kamiyama}}, \bibinfo {author} {\bibfnamefont
  {J.}~\bibnamefont {Campo}}, \bibinfo {author} {\bibfnamefont
  {K.}~\bibnamefont {Inoue}},\ and\ \bibinfo {author} {\bibfnamefont
  {J.}~\bibnamefont {Akimitsu}},\ }\bibfield  {title} {\bibinfo {title} {Long
  periodic helimagnetic ordering in Cr$M_3$S$_6$ ($M$= Nb and Ta)},\ }in\ \href
  {https://doi.org/10.1088/1742-6596/746/1/012061} {\emph {\bibinfo {booktitle}
  {J. Phys.: Conference Series}}},\ Vol.\ \bibinfo {volume} {746}\ (\bibinfo
  {organization} {IOP Publishing},\ \bibinfo {year} {2016})\ p.\ \bibinfo
  {pages} {012061}\BibitemShut {NoStop}%
\bibitem [{\citenamefont {Togawa}\ \emph {et~al.}(2012)\citenamefont {Togawa},
  \citenamefont {Koyama}, \citenamefont {Takayanagi}, \citenamefont {Mori},
  \citenamefont {Kousaka}, \citenamefont {Akimitsu}, \citenamefont {Nishihara},
  \citenamefont {Inoue}, \citenamefont {Ovchinnikov},\ and\ \citenamefont
  {Kishine}}]{togawa2012chiral}%
  \BibitemOpen
  \bibfield  {author} {\bibinfo {author} {\bibfnamefont {Y.}~\bibnamefont
  {Togawa}}, \bibinfo {author} {\bibfnamefont {T.}~\bibnamefont {Koyama}},
  \bibinfo {author} {\bibfnamefont {K.}~\bibnamefont {Takayanagi}}, \bibinfo
  {author} {\bibfnamefont {S.}~\bibnamefont {Mori}}, \bibinfo {author}
  {\bibfnamefont {Y.}~\bibnamefont {Kousaka}}, \bibinfo {author} {\bibfnamefont
  {J.}~\bibnamefont {Akimitsu}}, \bibinfo {author} {\bibfnamefont
  {S.}~\bibnamefont {Nishihara}}, \bibinfo {author} {\bibfnamefont
  {K.}~\bibnamefont {Inoue}}, \bibinfo {author} {\bibfnamefont
  {A.}~\bibnamefont {Ovchinnikov}},\ and\ \bibinfo {author} {\bibfnamefont
  {J.-i.}\ \bibnamefont {Kishine}},\ }\bibfield  {title} {\bibinfo {title}
  {Chiral magnetic soliton lattice on a chiral helimagnet},\ }\href
  {https://doi.org/10.1103/physrevlett.108.107202} {\bibfield  {journal}
  {\bibinfo  {journal} {Phys. Rev. Lett}\ }\textbf {\bibinfo {volume} {108}},\
  \bibinfo {pages} {107202} (\bibinfo {year} {2012})}\BibitemShut {NoStop}%
\bibitem [{\citenamefont {Zhang}\ \emph {et~al.}(2021)\citenamefont {Zhang},
  \citenamefont {Zhang}, \citenamefont {Liu}, \citenamefont {Zhang},
  \citenamefont {Yuan}, \citenamefont {Li}, \citenamefont {Wen}, \citenamefont
  {Jiang}, \citenamefont {Zhou}, \citenamefont {Lei} \emph
  {et~al.}}]{zhang2021chiral}%
  \BibitemOpen
  \bibfield  {author} {\bibinfo {author} {\bibfnamefont {C.}~\bibnamefont
  {Zhang}}, \bibinfo {author} {\bibfnamefont {J.}~\bibnamefont {Zhang}},
  \bibinfo {author} {\bibfnamefont {C.}~\bibnamefont {Liu}}, \bibinfo {author}
  {\bibfnamefont {S.}~\bibnamefont {Zhang}}, \bibinfo {author} {\bibfnamefont
  {Y.}~\bibnamefont {Yuan}}, \bibinfo {author} {\bibfnamefont {P.}~\bibnamefont
  {Li}}, \bibinfo {author} {\bibfnamefont {Y.}~\bibnamefont {Wen}}, \bibinfo
  {author} {\bibfnamefont {Z.}~\bibnamefont {Jiang}}, \bibinfo {author}
  {\bibfnamefont {B.}~\bibnamefont {Zhou}}, \bibinfo {author} {\bibfnamefont
  {Y.}~\bibnamefont {Lei}}, \emph {et~al.},\ }\bibfield  {title} {\bibinfo
  {title} {Chiral Helimagnetism and One-Dimensional Magnetic Solitons in a
  Cr-Intercalated Transition Metal Dichalcogenide},\ }\href
  {https://doi.org/10.1002/adma.202101131} {\bibfield  {journal} {\bibinfo
  {journal} {Adv. Mater.}\ }\textbf {\bibinfo {volume} {33}},\ \bibinfo {pages}
  {2101131} (\bibinfo {year} {2021})}\BibitemShut {NoStop}%
\bibitem [{\citenamefont {Hicken}\ \emph {et~al.}(2021)\citenamefont {Hicken},
  \citenamefont {Wilson}, \citenamefont {Franke}, \citenamefont {Huddart},
  \citenamefont {Hawkhead}, \citenamefont {Gomil{\v{s}}ek}, \citenamefont
  {Clark}, \citenamefont {Pratt}, \citenamefont {{\v{S}}tefan{\v{c}}i{\v{c}}},
  \citenamefont {Hall} \emph {et~al.}}]{hicken2021megahertz}%
  \BibitemOpen
  \bibfield  {author} {\bibinfo {author} {\bibfnamefont {T.}~\bibnamefont
  {Hicken}}, \bibinfo {author} {\bibfnamefont {M.}~\bibnamefont {Wilson}},
  \bibinfo {author} {\bibfnamefont {K.}~\bibnamefont {Franke}}, \bibinfo
  {author} {\bibfnamefont {B.}~\bibnamefont {Huddart}}, \bibinfo {author}
  {\bibfnamefont {Z.}~\bibnamefont {Hawkhead}}, \bibinfo {author}
  {\bibfnamefont {M.}~\bibnamefont {Gomil{\v{s}}ek}}, \bibinfo {author}
  {\bibfnamefont {S.}~\bibnamefont {Clark}}, \bibinfo {author} {\bibfnamefont
  {F.}~\bibnamefont {Pratt}}, \bibinfo {author} {\bibfnamefont
  {A.}~\bibnamefont {{\v{S}}tefan{\v{c}}i{\v{c}}}}, \bibinfo {author}
  {\bibfnamefont {A.}~\bibnamefont {Hall}}, \emph {et~al.},\ }\bibfield
  {title} {\bibinfo {title} {Megahertz dynamics in skyrmion systems probed with
  muon-spin relaxation},\ }\href {https://doi.org/10.1103/physrevb.103.024428}
  {\bibfield  {journal} {\bibinfo  {journal} {Phys. Rev. B.}\ }\textbf
  {\bibinfo {volume} {103}},\ \bibinfo {pages} {024428} (\bibinfo {year}
  {2021})}\BibitemShut {NoStop}%
\bibitem [{\citenamefont {Hicken}\ \emph
  {et~al.}(2022{\natexlab{b}})\citenamefont {Hicken}, \citenamefont {Wilson},
  \citenamefont {Holt}, \citenamefont {Khassanov}, \citenamefont {Lees},
  \citenamefont {Gupta}, \citenamefont {Das}, \citenamefont {Balakrishnan},\
  and\ \citenamefont {Lancaster}}]{hicken2022magnetism}%
  \BibitemOpen
  \bibfield  {author} {\bibinfo {author} {\bibfnamefont {T.}~\bibnamefont
  {Hicken}}, \bibinfo {author} {\bibfnamefont {M.}~\bibnamefont {Wilson}},
  \bibinfo {author} {\bibfnamefont {S.}~\bibnamefont {Holt}}, \bibinfo {author}
  {\bibfnamefont {R.}~\bibnamefont {Khassanov}}, \bibinfo {author}
  {\bibfnamefont {M.}~\bibnamefont {Lees}}, \bibinfo {author} {\bibfnamefont
  {R.}~\bibnamefont {Gupta}}, \bibinfo {author} {\bibfnamefont
  {D.}~\bibnamefont {Das}}, \bibinfo {author} {\bibfnamefont {G.}~\bibnamefont
  {Balakrishnan}},\ and\ \bibinfo {author} {\bibfnamefont {T.}~\bibnamefont
  {Lancaster}},\ }\bibfield  {title} {\bibinfo {title} {Magnetism in the
  N{\'e}el-skyrmion host GaV$_4$S$_8$ under pressure},\ }\href
  {https://doi.org/10.1103/PhysRevB.105.134414} {\bibfield  {journal} {\bibinfo
   {journal} {Phys. Rev. B.}\ }\textbf {\bibinfo {volume} {105}},\ \bibinfo
  {pages} {134414} (\bibinfo {year} {2022}{\natexlab{b}})}\BibitemShut
  {NoStop}%
\bibitem [{\citenamefont {Clark}\ \emph {et~al.}(2005)\citenamefont {Clark},
  \citenamefont {Segall}, \citenamefont {Pickard}, \citenamefont {Hasnip},
  \citenamefont {Probert}, \citenamefont {Refson},\ and\ \citenamefont
  {Payne}}]{clark2005first}%
  \BibitemOpen
  \bibfield  {author} {\bibinfo {author} {\bibfnamefont {S.~J.}\ \bibnamefont
  {Clark}}, \bibinfo {author} {\bibfnamefont {M.~D.}\ \bibnamefont {Segall}},
  \bibinfo {author} {\bibfnamefont {C.~J.}\ \bibnamefont {Pickard}}, \bibinfo
  {author} {\bibfnamefont {P.~J.}\ \bibnamefont {Hasnip}}, \bibinfo {author}
  {\bibfnamefont {M.~J.}\ \bibnamefont {Probert}}, \bibinfo {author}
  {\bibfnamefont {K.}~\bibnamefont {Refson}},\ and\ \bibinfo {author}
  {\bibfnamefont {M.}~\bibnamefont {Payne}},\ }\bibfield  {title} {\bibinfo
  {title} {First principles methods using {CASTEP}},\ }\href
  {https://doi.org/10.1524/zkri.220.5.567.65075} {\bibfield  {journal}
  {\bibinfo  {journal} {Z. Kristall.}\ }\textbf {\bibinfo {volume} {220}},\
  \bibinfo {pages} {567} (\bibinfo {year} {2005})}\BibitemShut {NoStop}%
\bibitem [{\citenamefont {Yates}\ \emph {et~al.}(2007)\citenamefont {Yates},
  \citenamefont {Wang}, \citenamefont {Vanderbilt},\ and\ \citenamefont
  {Souza}}]{yates2007spectral}%
  \BibitemOpen
  \bibfield  {author} {\bibinfo {author} {\bibfnamefont {J.~R.}\ \bibnamefont
  {Yates}}, \bibinfo {author} {\bibfnamefont {X.}~\bibnamefont {Wang}},
  \bibinfo {author} {\bibfnamefont {D.}~\bibnamefont {Vanderbilt}},\ and\
  \bibinfo {author} {\bibfnamefont {I.}~\bibnamefont {Souza}},\ }\bibfield
  {title} {\bibinfo {title} {Spectral and Fermi surface properties from Wannier
  interpolation},\ }\href {https://doi.org/10.1103/physrevb.75.195121}
  {\bibfield  {journal} {\bibinfo  {journal} {Phys. Rev. B.}\ }\textbf
  {\bibinfo {volume} {75}},\ \bibinfo {pages} {195121} (\bibinfo {year}
  {2007})}\BibitemShut {NoStop}%
\bibitem [{\citenamefont {Mulliken}(1955)}]{mulliken1955electronic}%
  \BibitemOpen
  \bibfield  {author} {\bibinfo {author} {\bibfnamefont {R.~S.}\ \bibnamefont
  {Mulliken}},\ }\bibfield  {title} {\bibinfo {title} {Electronic population
  analysis on LCAO--MO molecular wave functions. I},\ }\href
  {https://doi.org/10.1063/1.1740588} {\bibfield  {journal} {\bibinfo
  {journal} {J. Chem. Phys.}\ }\textbf {\bibinfo {volume} {23}},\ \bibinfo
  {pages} {1833} (\bibinfo {year} {1955})}\BibitemShut {NoStop}%
\bibitem [{sm()}]{sm}%
  \BibitemOpen
  \href@noop {} {\bibinfo {title} {See the Supplemental Material for further
  details that would allow one to reproduce our work, additional comparison to
  experimental measurements, DoS calculations for the materials in the series,
  and a complete set of Fermi surfaces.}}\BibitemShut {Stop}%
\bibitem [{\citenamefont {Van~den Berg}\ and\ \citenamefont
  {Cossee}(1968)}]{van1968structural}%
  \BibitemOpen
  \bibfield  {author} {\bibinfo {author} {\bibfnamefont {J.}~\bibnamefont
  {Van~den Berg}}\ and\ \bibinfo {author} {\bibfnamefont {P.}~\bibnamefont
  {Cossee}},\ }\bibfield  {title} {\bibinfo {title} {Structural aspects and
  magnetic behaviour of NbS$_2$ and TaS$_2$ containing extra metal atoms of the
  first transition series},\ }\href
  {https://doi.org/10.1016/S0020-1693(00)87012-7} {\bibfield  {journal}
  {\bibinfo  {journal} {Inorganica Chimica Acta}\ }\textbf {\bibinfo {volume}
  {2}},\ \bibinfo {pages} {143} (\bibinfo {year} {1968})}\BibitemShut {NoStop}%
\bibitem [{\citenamefont {Anzenhofer}\ \emph {et~al.}(1970)\citenamefont
  {Anzenhofer}, \citenamefont {Van Den~Berg}, \citenamefont {Cossee},\ and\
  \citenamefont {Helle}}]{anzenhofer1970crystal}%
  \BibitemOpen
  \bibfield  {author} {\bibinfo {author} {\bibfnamefont {K.}~\bibnamefont
  {Anzenhofer}}, \bibinfo {author} {\bibfnamefont {J.}~\bibnamefont {Van
  Den~Berg}}, \bibinfo {author} {\bibfnamefont {P.}~\bibnamefont {Cossee}},\
  and\ \bibinfo {author} {\bibfnamefont {J.}~\bibnamefont {Helle}},\ }\bibfield
   {title} {\bibinfo {title} {The crystal structure and magnetic
  susceptibilities of MnNb$_3$S$_6$, FeNb$_3$S$_6$, CoNb$_3$S$_6$ and
  NiNb$_3$S$_6$},\ }\href {https://doi.org/10.1016/0022-3697(70)90315-X}
  {\bibfield  {journal} {\bibinfo  {journal} {Journal of Physics and Chemistry
  of Solids}\ }\textbf {\bibinfo {volume} {31}},\ \bibinfo {pages} {1057}
  (\bibinfo {year} {1970})}\BibitemShut {NoStop}%
\bibitem [{\citenamefont {Hulliger}\ and\ \citenamefont
  {Pobitschka}(1970)}]{hulliger1970magnetic}%
  \BibitemOpen
  \bibfield  {author} {\bibinfo {author} {\bibfnamefont {F.}~\bibnamefont
  {Hulliger}}\ and\ \bibinfo {author} {\bibfnamefont {E.}~\bibnamefont
  {Pobitschka}},\ }\bibfield  {title} {\bibinfo {title} {On the magnetic
  behavior of new 2H-NbS$_2$-type derivatives},\ }\href
  {https://doi.org/10.1016/0022-4596(70)90001-0} {\bibfield  {journal}
  {\bibinfo  {journal} {Journal of Solid State Chemistry}\ }\textbf {\bibinfo
  {volume} {1}},\ \bibinfo {pages} {117} (\bibinfo {year} {1970})}\BibitemShut
  {NoStop}%
\bibitem [{\citenamefont {Friend}\ \emph {et~al.}(1977)\citenamefont {Friend},
  \citenamefont {Beal},\ and\ \citenamefont {Yoffe}}]{friend1977electrical}%
  \BibitemOpen
  \bibfield  {author} {\bibinfo {author} {\bibfnamefont {R.}~\bibnamefont
  {Friend}}, \bibinfo {author} {\bibfnamefont {A.}~\bibnamefont {Beal}},\ and\
  \bibinfo {author} {\bibfnamefont {A.}~\bibnamefont {Yoffe}},\ }\bibfield
  {title} {\bibinfo {title} {Electrical and magnetic properties of some first
  row transition metal intercalates of niobium disulphide},\ }\href
  {https://doi.org/10.1080/14786437708232952} {\bibfield  {journal} {\bibinfo
  {journal} {The Philosophical Magazine: A Journal of Theoretical Experimental
  and Applied Physics}\ }\textbf {\bibinfo {volume} {35}},\ \bibinfo {pages}
  {1269} (\bibinfo {year} {1977})}\BibitemShut {NoStop}%
\bibitem [{\citenamefont {Beal}(1979{\natexlab{a}})}]{beal1979first}%
  \BibitemOpen
  \bibfield  {author} {\bibinfo {author} {\bibfnamefont {A.}~\bibnamefont
  {Beal}},\ }\bibfield  {title} {\bibinfo {title} {The first row transition
  metal intercalation complexes of some metallic group VA transition metal
  dichalcogenides},\ }in\ \href {https://doi.org/10.1007/978-94-009-9415-7_5}
  {\emph {\bibinfo {booktitle} {Intercalated Layered Materials}}}\ (\bibinfo
  {publisher} {Springer},\ \bibinfo {year} {1979})\ pp.\ \bibinfo {pages}
  {251--305}\BibitemShut {NoStop}%
\bibitem [{\citenamefont {Parkin}\ and\ \citenamefont
  {Friend}(1980{\natexlab{c}})}]{parkin19803d}%
  \BibitemOpen
  \bibfield  {author} {\bibinfo {author} {\bibfnamefont {S.}~\bibnamefont
  {Parkin}}\ and\ \bibinfo {author} {\bibfnamefont {R.}~\bibnamefont
  {Friend}},\ }\bibfield  {title} {\bibinfo {title} {3d transition-metal
  intercalates of the niobium and tantalum dichalcogenides},\ }\href
  {https://doi.org/10.1080/13642818008245370} {\bibfield  {journal} {\bibinfo
  {journal} {Philosophical Magazine B}\ }\textbf {\bibinfo {volume} {41}},\
  \bibinfo {pages} {95} (\bibinfo {year} {1980}{\natexlab{c}})}\BibitemShut
  {NoStop}%
\bibitem [{\citenamefont {Gorochov}\ \emph {et~al.}(1981)\citenamefont
  {Gorochov}, \citenamefont {Blanc-soreau}, \citenamefont {Rouxel},
  \citenamefont {Imbert},\ and\ \citenamefont
  {Jehanno}}]{gorochov1981transport}%
  \BibitemOpen
  \bibfield  {author} {\bibinfo {author} {\bibfnamefont {O.}~\bibnamefont
  {Gorochov}}, \bibinfo {author} {\bibfnamefont {A.~L.}\ \bibnamefont
  {Blanc-soreau}}, \bibinfo {author} {\bibfnamefont {J.}~\bibnamefont
  {Rouxel}}, \bibinfo {author} {\bibfnamefont {P.}~\bibnamefont {Imbert}},\
  and\ \bibinfo {author} {\bibfnamefont {G.}~\bibnamefont {Jehanno}},\
  }\bibfield  {title} {\bibinfo {title} {Transport properties, magnetic
  susceptibility and M{\"o}ssbauer spectroscopy of Fe$_{0.25}$NbS$_2$ and
  Fe$_{0.33}$NbS$_2$},\ }\href {https://doi.org/10.1080/01418638108222164}
  {\bibfield  {journal} {\bibinfo  {journal} {Philosophical Magazine B}\
  }\textbf {\bibinfo {volume} {43}},\ \bibinfo {pages} {621} (\bibinfo {year}
  {1981})}\BibitemShut {NoStop}%
\bibitem [{\citenamefont {Haley}\ \emph {et~al.}(2020)\citenamefont {Haley},
  \citenamefont {Weber}, \citenamefont {Cookmeyer}, \citenamefont {Parker},
  \citenamefont {Maniv}, \citenamefont {Maksimovic}, \citenamefont {John},
  \citenamefont {Doyle}, \citenamefont {Maniv}, \citenamefont {Ramakrishna}
  \emph {et~al.}}]{haley2020half}%
  \BibitemOpen
  \bibfield  {author} {\bibinfo {author} {\bibfnamefont {S.~C.}\ \bibnamefont
  {Haley}}, \bibinfo {author} {\bibfnamefont {S.~F.}\ \bibnamefont {Weber}},
  \bibinfo {author} {\bibfnamefont {T.}~\bibnamefont {Cookmeyer}}, \bibinfo
  {author} {\bibfnamefont {D.~E.}\ \bibnamefont {Parker}}, \bibinfo {author}
  {\bibfnamefont {E.}~\bibnamefont {Maniv}}, \bibinfo {author} {\bibfnamefont
  {N.}~\bibnamefont {Maksimovic}}, \bibinfo {author} {\bibfnamefont
  {C.}~\bibnamefont {John}}, \bibinfo {author} {\bibfnamefont {S.}~\bibnamefont
  {Doyle}}, \bibinfo {author} {\bibfnamefont {A.}~\bibnamefont {Maniv}},
  \bibinfo {author} {\bibfnamefont {S.~K.}\ \bibnamefont {Ramakrishna}}, \emph
  {et~al.},\ }\bibfield  {title} {\bibinfo {title} {Half-magnetization plateau
  and the origin of threefold symmetry breaking in an electrically switchable
  triangular antiferromagnet},\ }\href
  {https://doi.org/10.1103/PhysRevResearch.2.043020} {\bibfield  {journal}
  {\bibinfo  {journal} {Physical Review Research}\ }\textbf {\bibinfo {volume}
  {2}},\ \bibinfo {pages} {043020} (\bibinfo {year} {2020})}\BibitemShut
  {NoStop}%
\bibitem [{\citenamefont {Beal}(1979{\natexlab{b}})}]{Beal1979}%
  \BibitemOpen
  \bibfield  {author} {\bibinfo {author} {\bibfnamefont {A.~R.}\ \bibnamefont
  {Beal}},\ }\bibinfo {title} {The First Row Transition Metal Intercalation
  Complexes of Some Metallic Group VA Transition Metal Dichalcogenides},\ in\
  \href {https://doi.org/10.1007/978-94-009-9415-7_5} {\emph {\bibinfo
  {booktitle} {Intercalated Layered Materials}}},\ \bibinfo {editor} {edited
  by\ \bibinfo {editor} {\bibfnamefont {F.}~\bibnamefont {L{\'e}vy}}}\
  (\bibinfo  {publisher} {Springer Netherlands},\ \bibinfo {address}
  {Dordrecht},\ \bibinfo {year} {1979})\ pp.\ \bibinfo {pages}
  {251--305}\BibitemShut {NoStop}%
\bibitem [{\citenamefont {G\"uller}\ \emph {et~al.}(2016)\citenamefont
  {G\"uller}, \citenamefont {Vildosola},\ and\ \citenamefont
  {Llois}}]{guller2016}%
  \BibitemOpen
  \bibfield  {author} {\bibinfo {author} {\bibfnamefont {F.}~\bibnamefont
  {G\"uller}}, \bibinfo {author} {\bibfnamefont {V.~L.}\ \bibnamefont
  {Vildosola}},\ and\ \bibinfo {author} {\bibfnamefont {A.~M.}\ \bibnamefont
  {Llois}},\ }\bibfield  {title} {\bibinfo {title} {Spin density wave
  instabilities in the ${\mathrm{NbS}}_{2}$ monolayer},\ }\href
  {https://doi.org/10.1103/PhysRevB.93.094434} {\bibfield  {journal} {\bibinfo
  {journal} {Phys. Rev. B}\ }\textbf {\bibinfo {volume} {93}},\ \bibinfo
  {pages} {094434} (\bibinfo {year} {2016})}\BibitemShut {NoStop}%
\end{thebibliography}%

\end{document}